\documentclass[journal=biomacromolecules,manuscript=article]{achemso}

\usepackage{chemformula} 
\usepackage[T1]{fontenc} 
\usepackage{blindtext}
\usepackage{amsmath,amsfonts,amssymb,amsthm,amscd,mathdots,bbm,graphicx,xcolor}
\usepackage{color}
\usepackage[latin1]{inputenc}
\usepackage[english]{babel}
\usepackage{amsmath,amssymb,amsfonts,amsthm,enumerate}
\usepackage{mathrsfs}
\usepackage[T1]{fontenc}
\usepackage{epsfig}
\usepackage{graphicx}

\usepackage{graphicx}
\usepackage{wrapfig}
\usepackage{graphicx}

\definecolor{refkeybis}{gray}{.65}
\definecolor{labelkeybis}{gray}{.65}
{\makeatletter
\def\SK@refcolor{\color{refkeybis}}%
\def\SK@labelcolor{\color{labelkeybis}}}

\newcommand{\bchi}{\mbox{\boldmath{$\chi$}}}

\newcommand{\R}{\mathbb{R}}

\newcommand{\iz}  {\mbox{\boldmath{$i$}}}

\newcommand{\Lz}  {\mbox{\boldmath{$L$}}}
\newcommand{\Bz}  {\mbox{\boldmath{$B$}}}
\newcommand{\Dz}  {\mbox{\boldmath{$D$}}}

\newcommand{\wz}  {\mbox{\boldmath{$w$}}}

\newcommand{\be}{\begin{equation}}
\newcommand{\ee}{\end{equation}}
\newcommand{\ba}{\begin{array}}
\newcommand{\ea}{\end{array}}

\renewcommand\theequation{\arabic{equation}}

\author{Giuseppe Florio}
\affiliation{Politecnico di Bari, Dipartimento di Ing. Civile, Ambientale, del Territorio, Edile e di Chimica, Via Re David 200, 70126 Bari, Italy}
\altaffiliation{INFN, Sezione di Bari, I-70126 Bari, Italy}
\author{Giuseppe 
Puglisi}
\affiliation{Politecnico di Bari, Dipartimento di Ing. Civile, Ambientale, del Territorio, Edile e di Chimica, Via Re David 200, 70126 Bari, Italy}
\email{giuseppe.puglisi@poliba.it}

\title[An \textsf{achemso} demo]
  {A predictive model for the thermomechanical overstretching transition of double stranded DNA}

\abbreviations{IR,NMR,UV}
\keywords{DNA,Double stranded molecules,Melting,Temperature effects,Phase transition}

\begin{document}

%
%
%
%
%

\begin{abstract}
  By extending the classical Peyrard-Bishop model, we are able to obtain a fully analytical description for the mechanical resistance of DNA under stretching at variable values of temperature, number of base pairs and intrachains and interchains bonds stiffness. In order to compare elasticity and temperature effects, we first analyze the system in the zero temperature mechanical limit, important to describe several experimental effects including possible hysteresis. We then analyze temperature effects in the framework of equilibrium statistical mechanics. In particular, we obtain an analytical expression for the temperature dependent melting force and unzipping assigned displacement in the thermodynamical limit, also depending on the relative stability of intra vs inter molecular bonds. Such results coincide with the purely mechanical model in the limit of zero temperature and with the denaturation temperature that we obtain with the classical transfer integral method. Based on our analytical results, explicit analysis of the phase diagrams and cooperativity parameters are obtained, where also discreteness effect can be accounted for. The obtained results are successfully applied in reproducing the thermomechanical experimental melting of DNA and the response of DNA hairpins. Due to its generality, the proposed  approach can be extended to other thermomechanically induced molecular melting phenomena. \end{abstract}


\section{Introduction}

\begin{figure}[t]
\resizebox{0.8\textwidth}{!}{\includegraphics{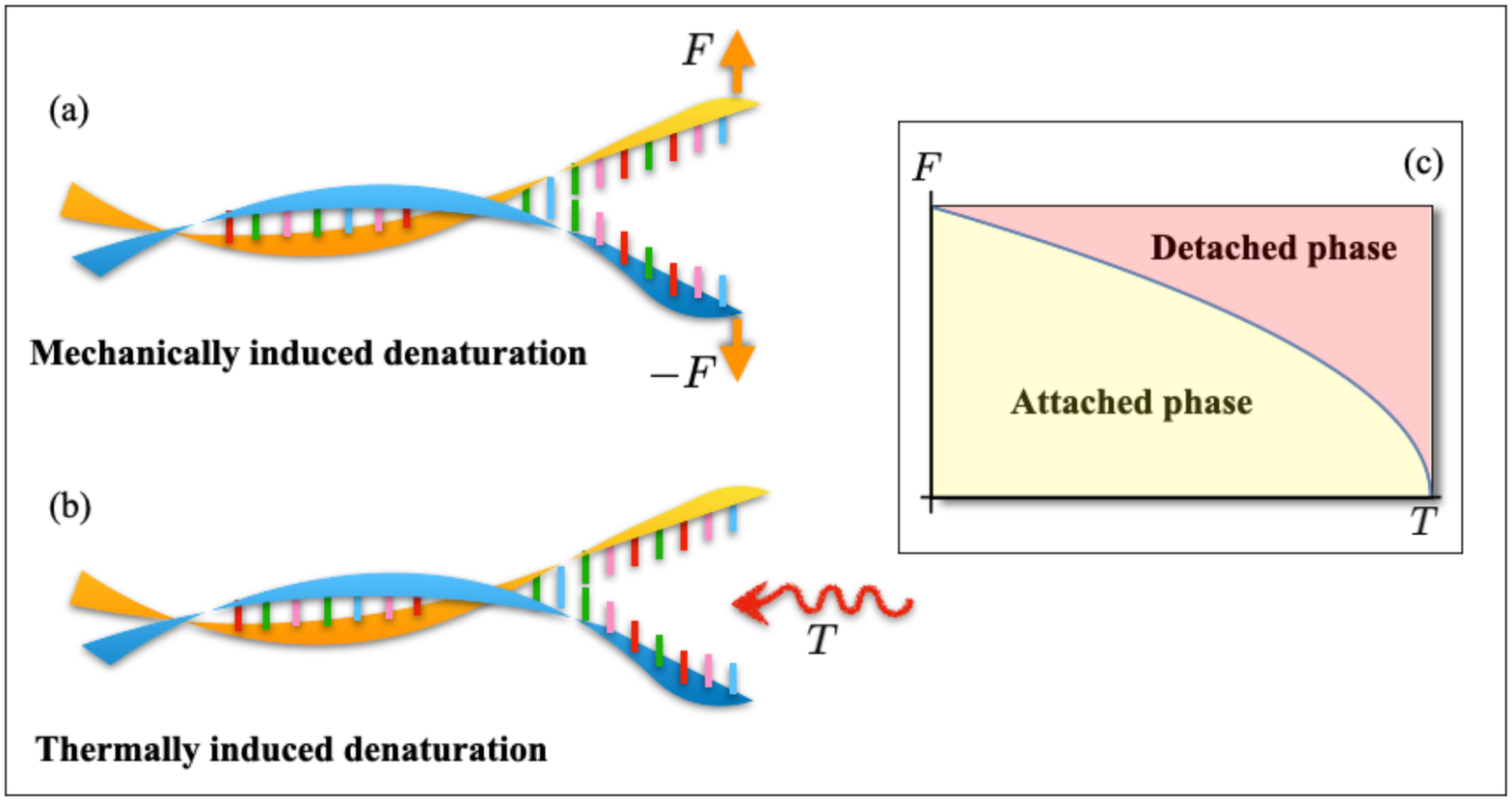}}\vspace{-0.5 cm}
\caption{Representation of the separation of two strands of DNA due to the action of a force (a) and induced by thermal effect (b). (c) Phase diagram of DNA denaturation.}
\label{dna-art} 
\end{figure}

\begin{figure}[t!]
\centering
$$\begin{array}{rr}
\includegraphics[height=10 cm]
{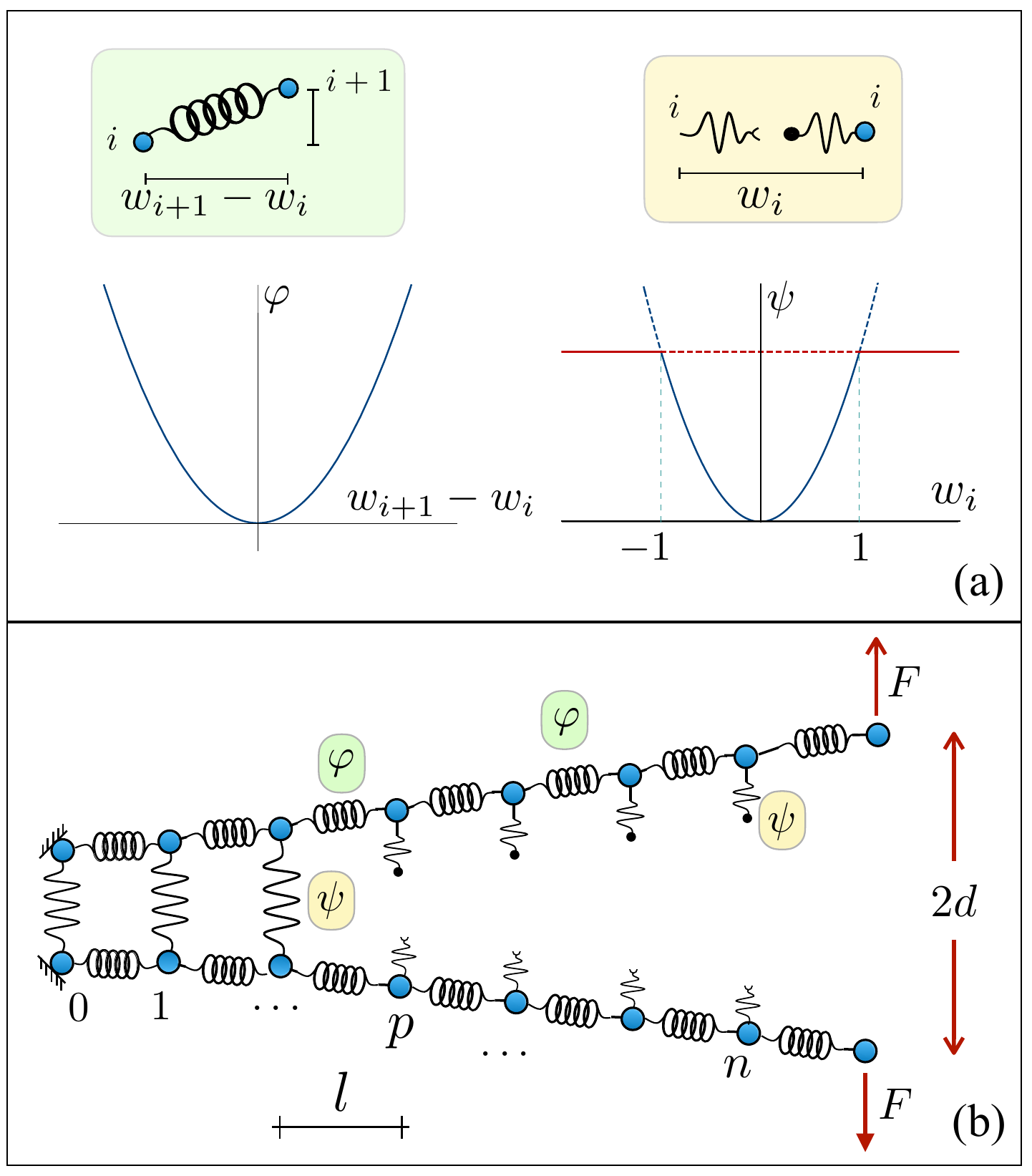}\\
\end{array}$$
\caption{(a) Scheme of the potential energy terms in the discrete model for DNA with the breakable link mimicking the behavior of the hydrogen bonds. The continuous line parabolic/constant is the single valued potential. Using the dashed lines we represent the two-values potential assumed when temperature effects are considered. (b) Scheme of the chain with breakable links and applied displacement to the final oscillator.}
\label{fig1} 
\end{figure}

After the pioneering  experimental work  in \cite{FMG}, mechanical molecular experiments  allowed increasingly sophisticated analyses of the energy landscape and mechanical stability at the single molecule scale. Stretching experiment of a single B-DNA molecule in \cite{SCB} revealed the presence of a force plateau that has been interpreted as a cooperative transition from a double-stranded (ds) configuration to a single-stranded (ss) one. Historically, the theoretical analysis of DNA has been based on the Peyrard-Bishop (PB)\cite {Pey} that allows to obtain numerical  and, under certain assumptions, analytical results. With the explicit aim of determining the influence of chain vs intra-strands bonds stiffnesses, temperature and discreteness effects for DNA and macromolecules with similar configurations (RNA, hairpins), here we use a PB type model for the analytical description of such melting transition. As in the case of the PB model, we neglect three-dimensional effects and inhomogeneity of the base-pairs sequences that may play an important role, but would hide the analytical clearness of the model.  Indeed, the main novelty of the paper is that, by substituting the classical Morse potential \cite {Pey}, describing breakable links, with a parabolic/constant energy (see Fig.~\ref{fig1}(a)) and using a spin-type approach (see \cite{FGP}),
we are able, both in the case of purely mechanical model (zero-temperature, Fig.~\ref{dna-art}(a)) and thermomechanically driven (Fig.~\ref{dna-art}(b)) melting processes, to obtain fully analytical results.  In particular, we are able to determine the crucial role of a main non-dimensional parameter $\nu$, measuring the relative stiffness of intra- vs inter-chains bonds on the melting transition behavior.
As we show, in accordance with known  experimental observations and reproducing typical AFM experiments,
in the case of assigned displacement orthogonal to the molecules (unzipping loading) we observe that the melting phenomenon is signaled by stress serrations with a nucleation stress higher than the propagation stress plateaux \cite{KR}. Similarly, we deduce the effect of the parameter $\nu$ and of temperature on the cooperativity of the transition. A detailed analysis of the elasticity properties on the cooperativity of the melting transition in the purely mechanical case and shear type loading for double stranded molecules has been recently analyzed by the autors in  \cite{BFGP}.

As we show, the behavior in the non zero temperature case, converges to the mechanical limit when temperature decreases. Moreover, the comparison with previous results in literature, either numerical or theoretical -- based on specific assumptions on the stiffness of the chain (e.g. continuum limit in \cite {Pey} and in \cite{Teo}  or \cite{TPM} in the extreme discretization limit) -- show the consistency of our results that do not require such specific assumptions. Finally we show the consistency of the proposed model regarding the denaturation temperature by comparing the results with those obtained by the transfer integral technique.

In summary, based on the recalled simplification of the model, we are able to fill all the gaps of the several long time analyses of the popular PB model, by fully characterizing its energy landscape in the zero temperature limit and by determining the behavior of the system independently on any assumption of the stiffness ratio $\nu$ between  the inter and intra-chains links.
Two main assumptions let us obtain these results. On one hand, we exploited the widely used single domain wall hypothesis (decomposing the chain into two complementary attached and detached part). We explicitly analyze such hypothesis that can be theoretically justified in the zipping loading cases when temperature does not approach the limit denaturation value. On the other, a second important simplification is that the analytic evaluation of the partition function is based on an assumption of multivalued configurational energy as explained in detail in the following. This assumption, firstly proposed in \cite{Seil}, has been numerically shown to be effective in \cite{FP}.

Finally, we obtain a clear phase diagram in the force-temperature plane  (Fig.~\ref{fig:deltacrit}(a)) and assigned displacement-temperature plane (Fig.~\ref{fig:deltacrit}(b)) determining
a stiffness-dependent analytic expression for the critical temperature. By substituting the known material parameters of the intra and inter-chains bonds we then show the possibility of quantitatively predicting the experimental behavior of the system with a very accurate result for all the range of temperature ranging from the zero limit to the denaturation value. As we show, the model can be extended for large systems also to quantitatively reproduce the experimental thermomechanical response of DNA hairpins. In both cases, we find a very good agreement between the theoretical predictions and the experimental results.

\vspace{0.2 cm}

\paragraph{Model --}\label{sec:model}

Following \cite{Pey}, we consider two chains of $n+1$ shear springs representing intrachains bonds  (see Fig.\ref{fig1}) and elasto-fragile extensional springs reproducing interchains base interactions. We assume symmetric behavior of the two strands, so that we can minimize the energy of a single chain connected to a fixed (symmetry) axis by breakable links representing inter-strands interactions. The simple but fundamental innovation with respect to classical analysis of the PB model is the assumption of a simplified form on the base interactions potential energy 
\begin{equation}\label{eq:potentialcut}
\Psi(u_i)=\frac{k_e l}{2} \left \{ \begin{array}{ll} (u_i/u_d)^2,  &\mbox{ if } |u_i/u_d|\leq 1 \\
1  &\mbox{ if } |u_i/u_d|>1 \end{array}\right .
\end{equation}
where $k_e>0$ is the elastic modulus of the unbroken links and $u_d$ is the debonding threshold, $u_i$, $i=0,1,..., n+1$, are the transverse displacements with respect to the symmetry axis of the chain, and we introduced the spring natural length $l=L/n$, $L$ being the total chain reference length. 
We may then introduce an internal (spin) variable $\chi_i$ with
\begin{equation}
\chi_i=\left \{ \begin{array}{ll} 0,  \mbox{ if } |u|\leq u_d, &\hspace{1 cm}\mbox{unbroken link,}\\
1,  \mbox{ if } |u|>u_d, &\hspace{1 cm}\mbox{broken link,}\end{array}\right .
\label{chi}
\end{equation}
thus describing each interchain link as a two-state element. 
The case of units undergoing transition between unbroken ad broken states  has been investigated with a similar approach in the cases of parallel links in \cite{Seil}
and, more recently, in \cite{ET, MPPT} and then applied to biological adhesion in \cite{PT2}.

As a result, the total potential energy of the breakable bonds is 
\begin{equation}\label{eq:phibr}
\Phi_{br}=\frac{k_e l}{2} \sum_{i=1}^n   \left[ (1-\chi_i) \left( \frac{u_i}{u_d}\right)^2+\chi_i  \right]
\end{equation}
whereas the strand total potential energy is 
\begin{equation}\label{eq:enelastic}
\Phi_{el}=\frac{k_t l}{2}  \sum_{i=0}^n \left (\frac{u_{i+1}-u_i }{l}\right)^2
\end{equation}
where $k_t>0$ is the shear constant.
Finally, the total potential energy of the system reads
\begin{equation}
\Phi=\Phi_{el}+\Phi_{br}. 
\end{equation}
We stress that, due to its definition in Eq.(\ref{eq:phibr}), $\Phi_{br}$ depends on the configuration vector $\bchi=\{\chi_1,\dots,\chi_n\}$.
To reproduce classical experiments on DNA molecules, we suppose that one side of the chain is fixed with $u_0=0$, whereas we assume that  a displacement $d$ is assigned to the last mass (so-called {\it hard device} condition, see Fig.~\ref{fig1}). These conditions represent a good approximation of hairpins unzipping when the extensional stiffness of the single stranded molecules significantly overcome the double-stranded one. 
More general boundary conditions taking care of the influence of the loading system can be assigned as proposed in \cite{FP} by considering another  energy term introducing the elasticity of the loading device or interacting molecule (such as RNA in the case of DNA transcription), but for the sake of simplicity we here neglect this additional term.

After introducing the rescaled displacements $w_i=u_i/u_d$ we can write the (non dimensional) total elastic energy as
\begin{equation}
n\, \phi= \frac{\Phi}{k_e l}=\frac{\nu^2}{2} \sum_{i=0}^n \left(w_{i+1}-w_i\right)^2
+\frac{1}{2} \sum_{i=1}^n \left [ (1-\chi_i) w_i^2+\chi_i \right ]. \label{hh}
\end{equation}
In Eq.(\ref{hh}) we introduced the main {\it non dimensional parameter} of the model
\begin{equation}
\nu=\sqrt{\frac{k_t}{k_e}}\, \frac{u_d}{l},\label{eq:nu}
\end{equation}
measuring the energy per unit (relative) displacement of the shear (inter-strand/covalent) bonds {\it vs} the energy per unit displacement for the breakable (intra-strands/non-covalent) bonds. 

As remarked in the introduction, previous analytical results in the literature are  based on specific assumptions such as continuum limit in \cite {Pey} and in \cite{Teo}  or \cite{TPM} in the extreme discretization limit (corresponding to $\frac{\nu^2}{n}\rightarrow 0$ and $ \frac{\nu^2}{n}\rightarrow \infty$, respectively). Here we do not need any of these assumption so that both limits result as particular cases of our analysis.

\section{Mechanical limit}\label{sec:zerotemp}
Consider first the mechanical (zero-temperature) limit describing the behavior when entropic  terms can be neglected as compared to the internal energy. In this case the observed configurations correspond to the minima of the energy in Eq. \eqref{hh}.

\paragraph{Equilibrium solutions --}
In order to take care of the constraint of assigned displacement, we introduce the Lagrange multiplier $f$ coupled to the assigned $\delta=d/u_d$, where $d$ is the imposed end-point displacement (see Fig.~\ref{fig1}), and minimize the function
\begin{equation}\label{eq:ng}
n\, g=n\, \phi-f \delta.
\end{equation}
We notice that the parameter $f$ coupled to $\delta$ represents the (rescaled, non-dimensional) force acting on the system and its relation with the effective measured force $F$ is
\begin{equation}
F=\frac{k_e l}{u_d}f.\label{eq:forceexp}
\end{equation}
It is possible to see (Supplementary Information, SI) that the equilibrium solutions $w_i$ are monotonically increasing for $\delta>0$, {\it i.e.} $w_{i+1}\ge w_{i}$ for $i=0\dots,n-1$.
As a result, by using Eq.\eqref{chi}, all the stable equilibrium solutions are characterized by an initial connected segment of $p$ attached links and $n-p$ detached links, so-called {\it single domain wall} (SDW) solutions. It is important to remark that this condition in general is {\it not} verified when the effects of the temperature $T$ are taken into account. On the other hand, the efficacy of this approximation in the case of $T>0$ at realistic values of the temperature has been numerically shown for protein unfolding experiments in \cite{FP}. 

An explicit calculation (SI) allows us to evaluate the equilibrium values of $w_i$. In particular, given a fixed value of unbroken links $p$, the following total force-displacement relation holds
  \begin{equation}\label{eq:force-displ}
  w_{n+1}=\delta=\frac{f}{k(p)},
   \end{equation}
where we introduced the global stiffness (dependent on the configuration $p$)
\begin{equation}
k(p)=\frac{\nu^2}{\gamma(p)},  \,\,\, \,\,\,\, p=0,\dots,n, \label{kk}
\end{equation}
with
\begin{equation}\begin{array}{l}
\gamma(p)=n-p+\alpha(p),\vspace{0.2 cm}\\ \alpha(p)=\left \{1-\frac{\sinh{(p \lambda)}}{\sinh{[(p+1)\lambda]}}\right\}^{-1}, \vspace{0.2 cm}\\ \cosh{\lambda}=1+\frac{1}{2 \nu^2}.\end{array}
\end{equation}
The energy corresponding to each solution (at fixed $p$) reads
\begin{equation}
 n\,  \phi=\frac{  k(p) }{2}\delta^2+\frac{n-p}{2},
\label{equilenp} \end{equation}
where the first term represents the elastic energy of the equilibrium solution and the second term describes the (dissipated) unbonding energy.

When temperature effects are neglected, following \cite{MPPT, PT}, we can consider two limit cases. In the first scenario we assume that the system is able to overcome {\it any} energy barrier so that it is always able to relax to the global minimum of the total potential energy ({\it Maxwell convention}). In the second case the system is able to overcome {\it no} energy barrier and, therefore, it stays in a local minimum following an equilibrium branch of the energy in Eq.(\ref{equilenp})  ({\it Maximum delay convention}). \vspace{0.3 cm}

\paragraph{Global energy minima}\label{sec:global}
To determine the global minima we notice that the energy equilibrium branches in Eq.\eqref{equilenp} (each identified by the value of $p$) are convex. We indicate by $\delta_{Max}(p),\,p=0,...,n-1$ 
the positive intersection displacement of the
 $p$ and $p+1$ branches.

\begin{figure}[b!]
\includegraphics[height=6 cm] {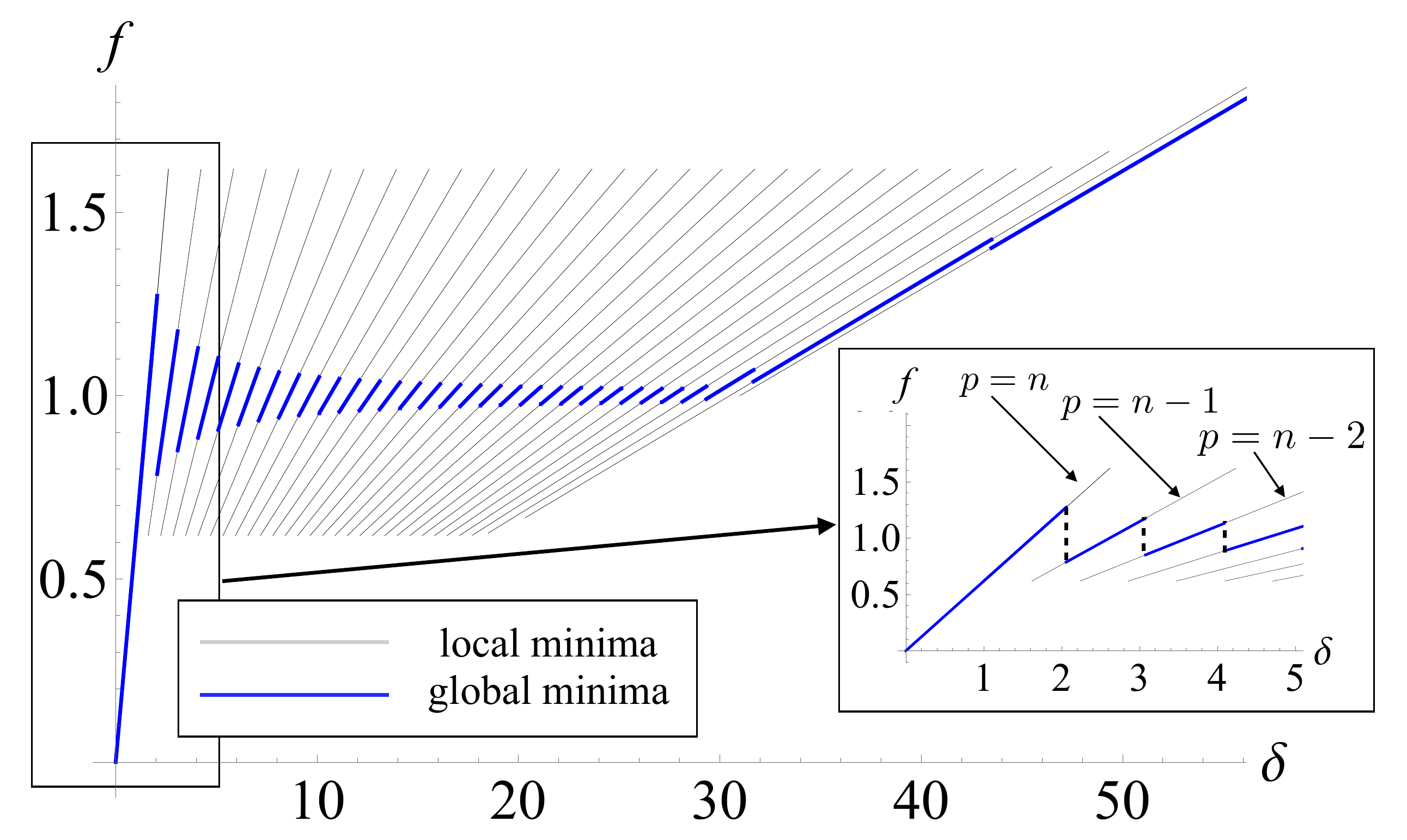}
\caption{Force-displacement relation under the Maxwell convention. Thin black lines represent local minima, thick blue line global minima. Here $n=30$ and $\nu=1$. 
  \label{ff3}}
\end{figure}

As a result we can find (see SI and, in particular, Eq.\eqref{eq:deltamax} for the details) the $f-\delta$ relation as represented with bold lines in Fig.\ref{ff3}. The melting transition is obtained by a serrated  plateaux with the decohesion front coherently propagating accordingly to the classical zipper hypothesis \cite{Kittel}. In particular, the force-displacement relation is given by Eq.(\ref{eq:force-displ}), following the conditions prescribed in Eq.(\ref{eq:deltagm}). We deduce that, according to the Maxwell convention, the system behaves elastically with $f=k(n)\delta$ until $\delta=\delta_{Max}(n-1)$. After this value we observe a sequence of successive jumps between neighbor branches (each denoted by the value of unbroken links $p$) with the value of the force $f$ in a $p$-dependent interval ($p>0$) as $f\in[f_1(p),f_2(p)]$ where $f_1(p)= k(p)\delta_{Max}(p),\, f_2(p)=k(p)\delta_{Max}(p-1)$.

\begin{figure}[h]
\includegraphics[height=6 cm]{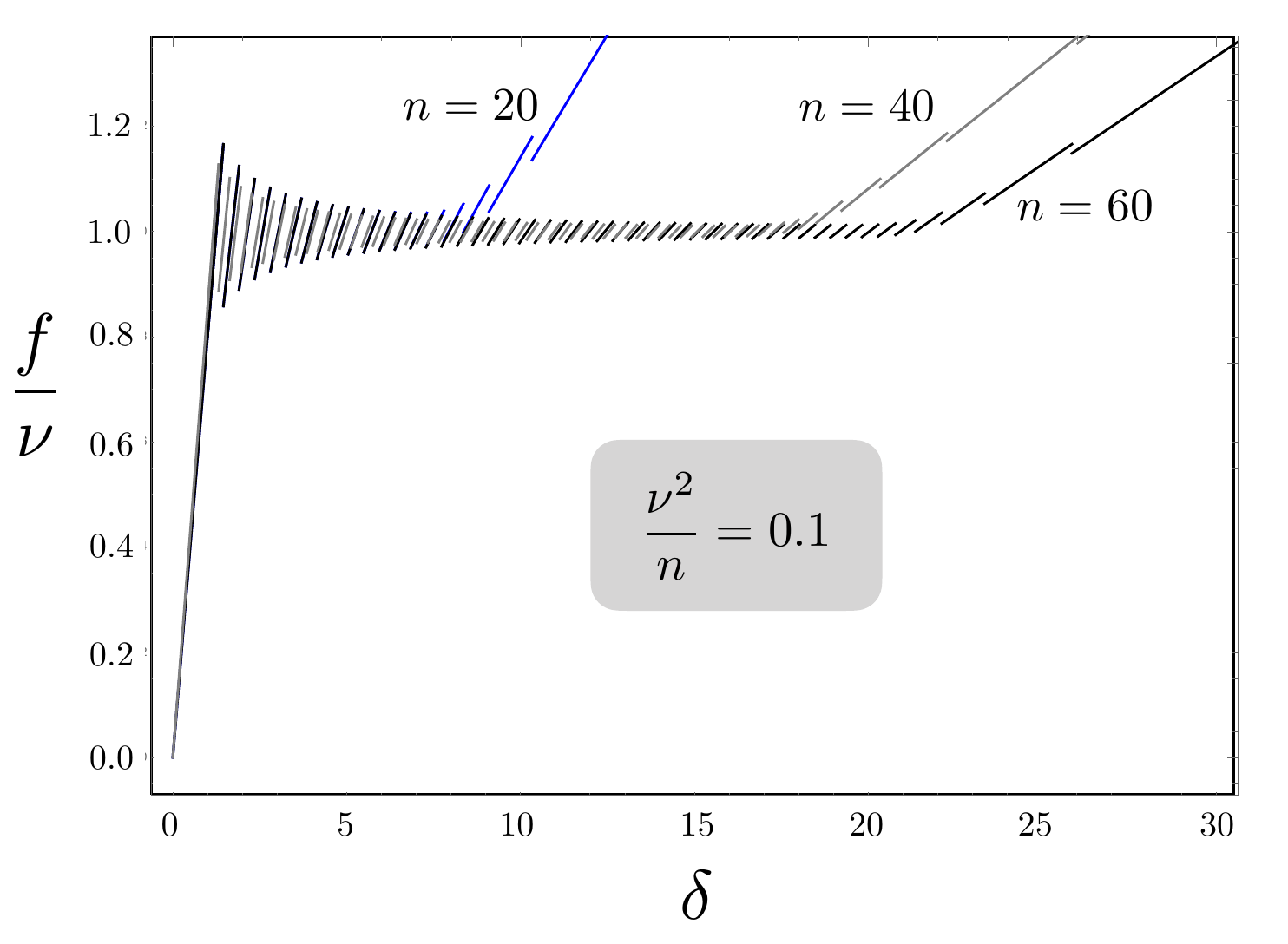}
\caption{ Dependence, under the global energy minimization hypothesis, of the melting transition behavior for a system with with variable discreteness parameter $n$ and $\nu$. \label{f4a}}
\end{figure}

It is possible to see that, initially, $f_2(p)$ decreases as $p$ decreases. This means that after the first link has been broken, the force necessary to break the following one is smaller, i.e. the decohesion process is more easily realized. On the other hand, due to a hardening effect induced by the constraint $w_0=0$, we observe a final elastic branch corresponding to $p=0$.  In the unconstrained case, the system would simply exhibits a softening behavior up to full melting. We also observe that the successive unbonding of the interchains links leading to serrations phenomena represents a well known experimental effect  (see {\it e.g.} \cite{KR}).  In Fig.\ref{f4a} we also include some details related to the influence of $\nu$ and of discreteness scale on the force-displacement relation. In the figure we represented the equilibrium solutions at different value of $n$ and $\nu$ with constant $\nu^2/n$. For values of $n$ large enough, this corresponds to fix the stiffness of the partially detached configurations (see \eqref{eq:ftl}). Moreover, we represent the rescaled force $f/\nu$ so that, again for large $n$, the plateaux is fixed (see \eqref{spsp}).
In particular the figure shows the important effect that the cooperativity of the melting transition increases as $\nu$ grows and $n$ decreases. 
This behavior is consistent with previous results in \cite{PT,MPPT}.

\vspace{0.3 cm}

\paragraph{Local energy minima}\label{sec:local}
If the system is not able to overcome any energy barrier, it follows a given configuration, fixed by $p$, until it becomes unstable. Thus, if we increase the displacement starting from the fully attached state ($p=n$), the system follows the first equilibrium branch until the last interchain bond attains its limit value $w_n=1$ and the force reaches the value $f_{\mbox{\tiny{\it M}}}(p)$ (see Eq.\eqref{fmM}-(\ref{deltamM}) in  SI). 
\begin{figure}[t]
\resizebox{0.85\textwidth}{!}{\includegraphics{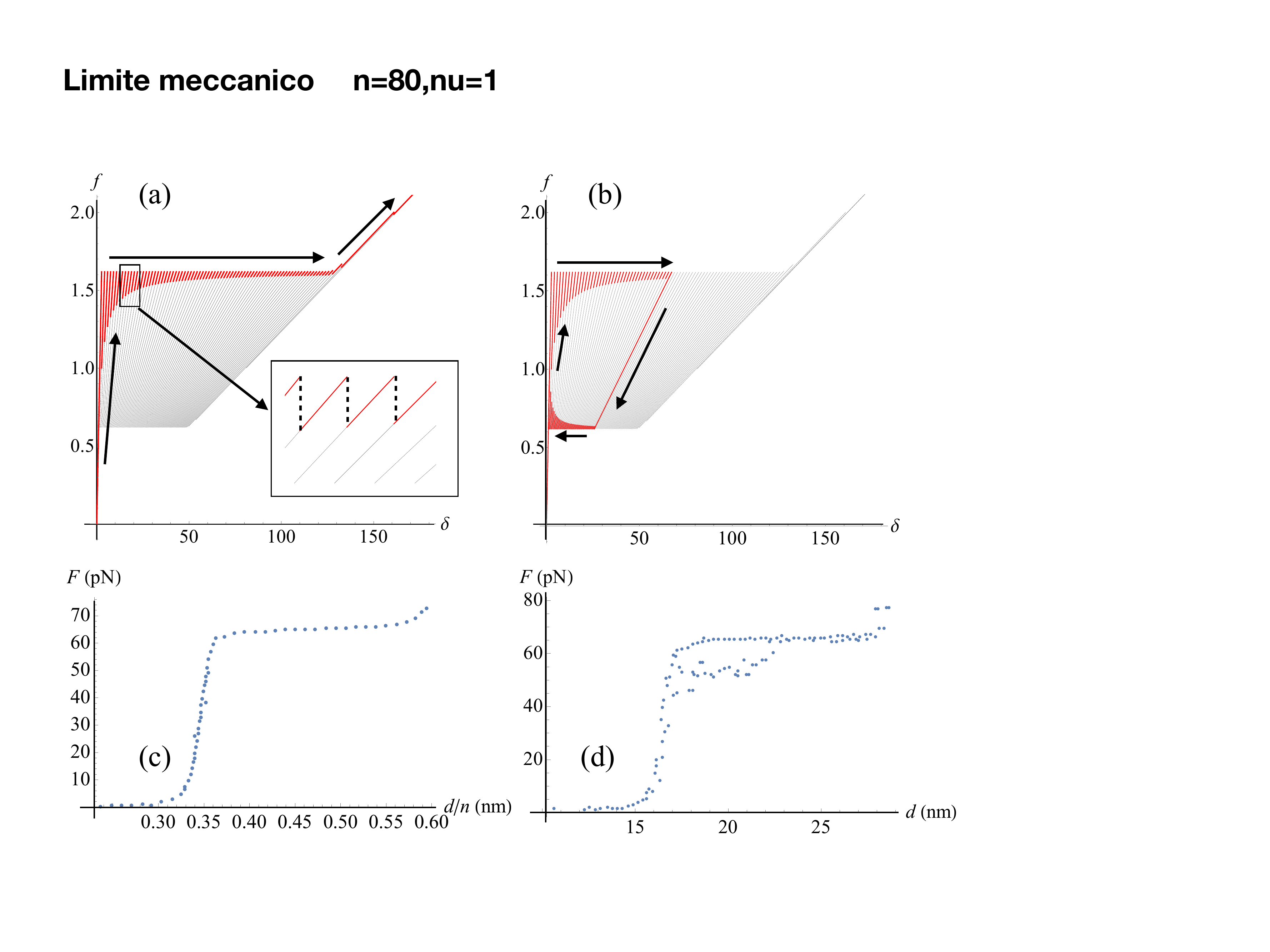}}
\caption{Equilibrium solutions for a system with $n=80$ and $\nu=1$ under the maximum delay convention. (a) Monotonic and (b) cyclic loading. Experimental force-displacement
diagram for (c) monotonic (extracted from \cite{RB}) and for (d) cyclic loading
(extracted from \cite{SCB}).\label{f3}}
\end{figure}
Beyond this threshold, we observe a jump from one branch to the following one with a corresponding measured force $f_{\mbox{\tiny{\it m}}}(n-1)$ (Eq.\eqref{fmM}-(\ref{deltamM})). 
For increasing displacement we obtain again a successive coherent propagation of the debonding front as represented in Fig.~\ref{f3} (a)-(b) where we show both the cases of monotonic and cyclic behavior. In the cyclic case the system describes an hysteretic path. In particular, if the load is reverted after $n-p$ bonds are broken, a reversible rebonding is allowed, the 
links rebond one at a time at a force $f_{\mbox{\tiny{\it m}}}(p)$ in the inverse order of the debonding process with  a coherent propagation of the (rebonding) front. These results for monotonically increasing and cyclic loading of the DNA molecules are in agreement  with the experimental behavior reproduced in Fig.\ref{f3}(c)-(d).

It is worth noticing that, differently from previous results, to the knowledge of the authors only numerical solutions for the equilibrium problem for PB models of DNA have been previously proposed (see {\it e.g.} \cite{TPM}).  

\section{Temperature effects}\label{sec:finitetemp}

In order to include the fundamental temperature effects into the DNA melting behavior it is possible to resort to the methods of Equilibrium Statistical Mechanics. As for the mechanical (zero-temperature) case, we consider the scenario where the displacement of the last oscillator is assigned, i.e. $w_{n+1}=\delta$. In terms of Statistical Mechanics, this choice corresponds to consider the so-called Helmholtz ensemble. The relations between macrosopic quantities can be obtained from the partition function defined as
\begin{eqnarray}\label{eq:partition}
\mathcal{Z}&=&\sum_{\{\chi_i\},\, i=1}^n  \int_{\R^{n}}  e^{-\beta n \phi(\wz,\delta)}\,\, d \wz 
\end{eqnarray}
evaluated  with $w_0=0$ and $w_{n+1}=\delta$. Here we introduced the $n$-components vector $\wz=\{w_1,w_2,\dots,w_n\}$. Moreover, we have used the symbol $\{\chi_i\}$ to represent the summation over all phase configurations with $\chi_i=\{0,1\}$. Finally, according with the non dimensionalization  in \eqref{hh}, we adopted the rescaled (non dimensional) inverse temperature $\beta$ as
\begin{equation}
\beta=\frac{k_e l}{k_B T}
\end{equation}
where $k_B$ is the Boltzman constant and $T$ the absolute temperature.
An explicit calculation of $\mathcal{Z}$ (see SI) allows to evaluate the free energy and the expected value of the coupled force
\begin{eqnarray}
\mathcal{F}(\beta, \delta)&=&-\frac{1}{\beta}\ln \mathcal{Z}(\beta, \delta),\quad \bar{f}(\beta, \delta)=\frac{\partial \mathcal{F}}{\partial\delta}
\label{eq:ansol}.
\end{eqnarray}
In order to simplify the analysis of the force-displacement relation and obtain fully analytical results we can use the SDW solutions introduced in the mechanical (zero temperature) case. This represents the well known zipper hypothesis \cite{Kittel} that has been energetically justified in the previous case of purely mechanical system, whereas it represents only an approximation when temperature effects are included. At the same time,  this approach has been numerically justified in the temperature range of real systems in \cite{FP}. In this case, the summation over all possible phase configurations in Eq.(\ref{eq:partition}) is substituted by a summation over the index $p$ denoting the number of unbroken links. This approximation is particularly useful for two reasons. On one hand, it makes the analytical expression more transparent, allowing, in principle, to evaluate the different statistical contributions of SDW (largely dominant in the range of temperature experienced in experiments) and non-SDW configurations. On the other hand, this approximation allows to obtain the fundamental relation between temperature and force in the thermodynamic limit. 

\begin{figure}[t!]
\centering
\resizebox{0.85\textwidth}{!}{\includegraphics{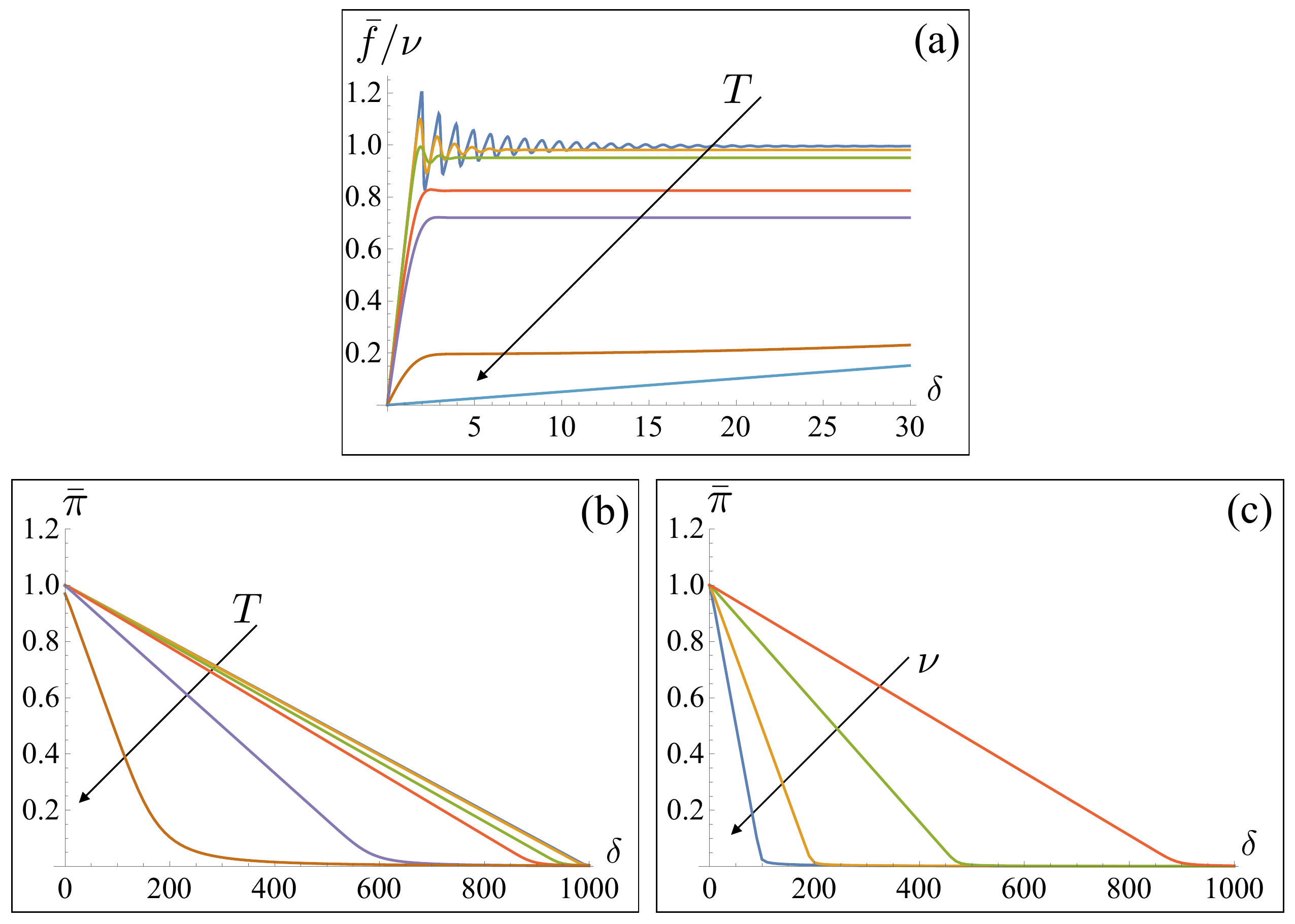}}
\caption{(a) Force-displacement curves for different values of temperature. Parameters: $n=200$, $\beta=100,25,10,3,2,1,0.5$. Dependence of $\bar{\pi}$ on $\delta$ using $T$ (b), $\nu$ (c)  $n=50,300,1000$, $\nu=1$, $\beta=5$. (d) $n=1000$, $\nu=10,5,2,1$, $\beta=5$.}
\label{fig:Fdelta} 
\end{figure}

The important role of temperature in mechanically driven melting transition is described  in Fig. \ref{fig:Fdelta}(a). Observe that, due to temperature effects, the horizontal serrated plateaux are substituted by smoother plateaux.
 We notice that in the limit of small temperature (large $\beta$, mechanical limit), the curve oscillates around the value of $\nu$ as obtained in the zero temperature limit (see Eq.\eqref{spsp}). On the other hand, for larger temperatures the folding process begins at lower values of the assigned displacement and resulting force, due to entropic effects. Moreover, the serrated path is smoothed. 
 
We may observe that the main effect of the proposed model is that the melting transition begins at significantly lower forces as temperature grows.  Moreover, it is interesting to analyze the temperature dependence of the expectation value of the fraction of unbroken bonds as the displacement is increased, giving a measure of the cooperativity of the denaturation phenomenon. In particular, we define $\bar{\pi}=\langle \pi \rangle=\langle p/n \rangle$ that can be evaluated starting from the canonical partition function $\mathcal{Z}$ (see SI).
In Fig.\ref{fig:Fdelta}(b)-(c) we report the dependence of this quantity on $T$ and $\nu$.
Again we may observe that the cooperativity grows as $\nu$ and $T$ are increased.
 As the figure shows, cooperativity grows also as the temperature decreases and the melting process is activated from lower forces as the temperature grows.

\vspace{0.2 cm}

\paragraph{Thermodynamic limit}
The behavior of macromolecules with a large number of bases $n$ can be of particular interest in the case of temperature effects because it allows us deducing fully analytical relations between the melting force and the temperature. Thus, we consider  the thermodynamic limit when $n$ (and $L=nl$) goes to infinity (with $l$ fixed). Indeed, by keeping the SDW approximation it is possible to obtain (SI) the analytic relation
\begin{eqnarray}\label{eq:forcetd}
\bar{f}&\simeq&\nu\sqrt{1-\frac{T}{T_c}},
\end{eqnarray}
where we have defined the critical temperature (independent on the applied displacement and conjugated force) 
\begin{eqnarray}\label{eq:Tc}
T_c=\frac{k_e \,l}{k_B\,\lambda}\end{eqnarray} 
such that the force threshold for melting transition decreases to zero and the DNA molecule spontaneously melts. 
We notice that  for $T=0$ we find $\bar{f}=\nu$ coherently with \eqref{spsp}.

We point out that the results in the thermodynamic limit can be obtained by the Transfer Integral (TI) method \cite{Pey}, but only in the limit of large values of $\nu^2$ as proposed also  in \cite{Singh} where the authors obtain a square root relation of the type in \eqref{eq:Tc}. In SI we deduce, based on the TI technique, the critical temperature $T_c^{(TI)}=\frac{\sqrt{k_t\,k_e}}{k_B}u_d$ for our model coherent with  (\ref{eq:Tc}) in the regime $\nu^2\gg1$. From this point of view, it may be of interest to compare the obtained results with those deduced in the literature based on the adoption of the Morse potential \cite{Pey,DTP,Teo,TPM}, defined as $V_{M}(y)=D\,(1-e^{-a y})^2$, with $D$ and $a$ representing the depth and the width of the potential well, respectively. Observe that our interchain interaction energy is symmetric with the possibility of detachment both for positive and negative displacements. Specifically, we considered the superposition of a quadratic energy term and a constant term (corresponding to the breaking of a bond). On the other hand, the Morse potential is asymmetric and in the TI approach this corresponds to the possibility of obtaining the critical temperature based on the existence criterion of bound states in the energy spectrum \cite{landau}. Moreover, to refer to previously recalled paper notation in the literature, we remark that the authors introduced the non dimensional parameter $R=Da^2/K$ (with $K$ the elastic constant in the Peyrard-Bishop model) such that $R\gg1$ and $R\ll1$ correspond to the discrete\cite{TPM} and continuous \cite{DTP} regimes, respectively. We remark that the analytical solution of the formal Schr$\ddot{\rm o}$edinger equation associated to the TI approach is valid if the condition $D\ll K_B T\ll D/R$ holds. On the other hand, the typical choice of material parameters in DNA models (see later for a comparison with experimental data) correspond to the former case (where typically $R\simeq O(10)$).  Making a comparison with the presented model, we notice that the direct method for the evaluation of the partition function based on the SDW approximation and the summation over phase (spin) variables is valid for {\it all} values of the parameter $\nu$, thus enlarging the range of possible application of the model and providing evidence of the more general nature of the methods used in this paper as compared with previous approaches that all requires  hypotheses on $\nu$. Roughly speaking, we may summarize that the model proposed in \cite{TPM} is able to describe the initial linear dependence for small value of $\nu$ and temperature, where the solutions approximate the purely mechanical solutions at zero temperature. On the other hand, the TI approach is appropriate to describe the regime of large values of $\nu$ and $T$ approximating the critical value. The results obtained from the model here proposed are independent on the value of $\nu$ and coherent with previously deduced approximate results.

\begin{figure}[t!]
\centering
{\includegraphics[height=5.15 cm]{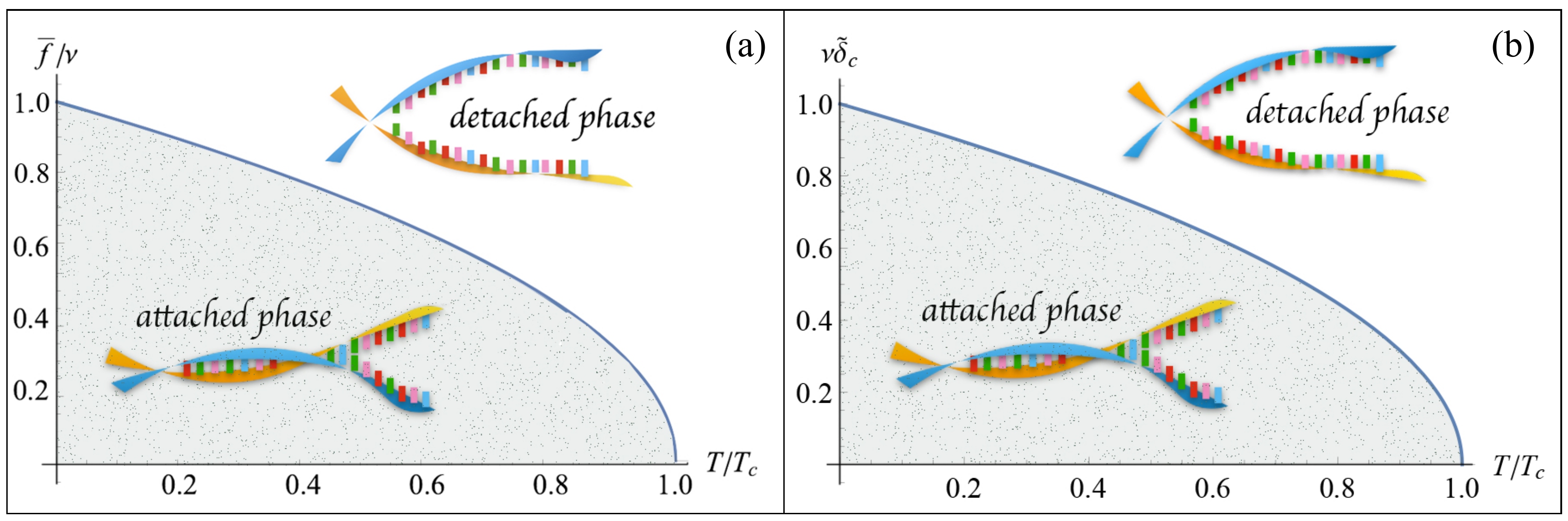}}
\caption{Phase diagram in the temperature-melting force (a)   temperature-crititical displacement (b) plane.}
\label{fig:deltacrit} 
\end{figure}
 
 \subsection{DNA melting as a phase transition}
 
 We here show that the melting of the DNA can be treated in the framework of phase transition theory.
It is interesting to point out that (see \cite{RB}) Eq.(\ref{eq:forcetd}) can be used to derive the phase diagram shown in Fig.\ref{fig:deltacrit}(a) separating the force-temperature plane
in regions with fully attached solutions and fully detached solutions, respectively. We can also derive a phase diagram for the applied displacement and temperature as shown in Fig.~\ref{fig:deltacrit}(b). Indeed, we may introduce $\tilde{\delta}_c=\delta_c/n$ as the (temperature and elasticity dependent) applied strain measure needed to obtain a complete denaturation of the chain when the expected value of the phase fraction of attached bonds $\pi=p/n$ attains the zero value. An explicit calculation in the thermodynamical limit (see SI) shows that
\begin{equation}
\tilde{\delta}_c=\frac{1}{\nu}\sqrt{1-\frac{T}{T_c}}.
\end{equation}

\begin{figure}[h!]
\centering
{\includegraphics[height=11 cm]{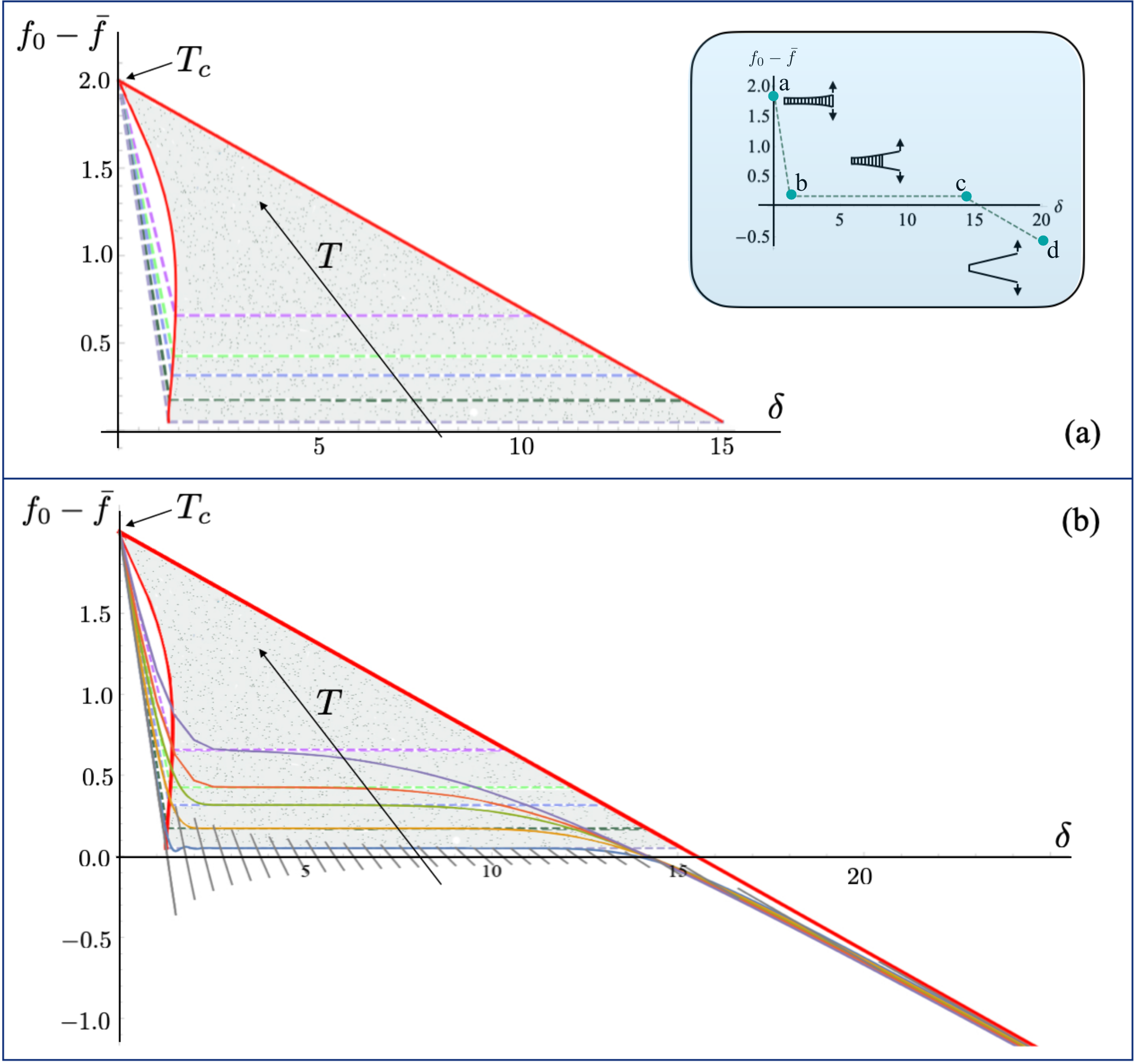}}
\caption{DNA melting as a phase transition. Grey regions represent partially attached configurations. (a): dependence of the difference $f_0-\bar{f}$ on $\delta$  in the thermodynamical limit with $f_0=\nu$ (see the text for the description); inset: correspondence of the curve branches with the configuration of the chain. (b): behavior for a limited DNA chain with $n=30$ superimposed to the ideal behavior in the thermodynamic limit in the thermodynamical limit; the behavior of the  finite size system in the mechanical limit is reported by gray equilibrium branches. In both panels we used $\nu=2$ and $\beta= 0.9, 1.3, 1.7, 3, 10$. The critical temperature $T_c$ corresponds to $\beta_c=\lambda(2)\sim 0.495$. In the inset of (a) we reported the behavior for $\beta=3$. }
\label{fig:deltacritb} 
\end{figure}

The transition behavior can be described in more details if we consider also the behavior at different temperatures in the force-displacement plane as represented in Fig.\ref{fig:deltacritb}. In panel (a) we show the dependence of the difference $f_0-\bar{f}$  on the applied rescaled displacement $\delta$ behavior in the thermodynamical limit (with $f_0=\nu$ the value of the force plateau in the mechanical limit). As shown in the inset, by increasing $\delta$, the system behaves elastically up to a critical value of displacement and force $\bar{f}$ (point a), both decreasing as the temperature increases, when the system starts to melt. In terms of a phase transition, this threshold (whose interpolation for the different curves is represented by a thick red line on the left side of the plot) represents the starting value of the passage from the single phase (fully attached) solution to the coexistence of two phases configurations with the system decomposed in an attached and in a detached domain. This region corresponds to a force plateaux (from point b to point c in the inset) with decreasing extension as the temperature increases. Thus, the transition cooperativity, in the considered case of assigned displacement, increases as temperature grows. As $\delta$ is increased, a second threshold is attained (point c) corresponding to reach the second single phase (fully detached) configuration, with the complete melting of the DNA chain. The system then follows the fully detached branch (from point c to point d in the figure, red thick line on the right of the main plot). The corresponding elasticity is due to the chain stiffness and the assumed boundary conditions. Full denaturation would be attained without the considered constraint $w_0=0$. In panel (b) discreteness effects are evidenced. In particular, the behavior of curves in the thermodynamical limit is compared to a short DNA double stranded molecule with $n=30$ bases. Observe that, due to the finite $n$, the transition corresponds to a sloped plateaux with an `anticipated' transition induced by the entropic terms, energetically favoring two-phases configurations.

\vspace{0.2 cm} 

\subsection{Comparison with experimental results}
We now compare the analytical results of previous section, regarding temperature effects,  with experimental data reported in \cite{rouzina}. 
Taking into account Eqs.(\ref{eq:nu}) and (\ref{eq:forceexp}), we find the expression of the (temperature-dependent) expectation value of the physical force in terms of the material parameters of the model:
\begin{equation}
\bar{F}=\sqrt{ k_e\,k_t\left(1-\frac{T}{T_c}\right ) }.
\label{eq:fvst}
\end{equation}

\begin{figure}
\centering
{\includegraphics[height=6 cm]{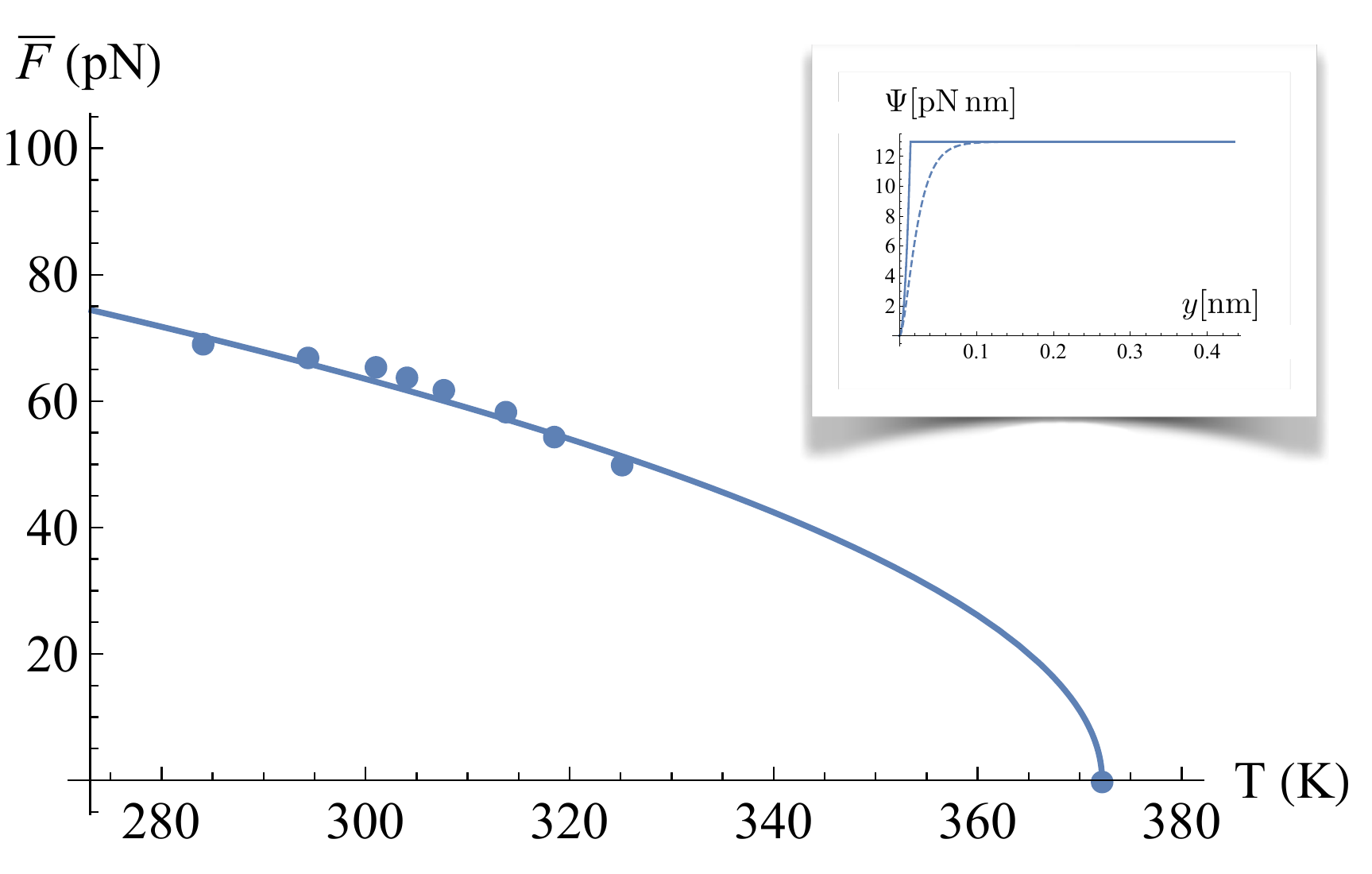}}
\caption{Comparison of experimental results with the theoretical prediction of the denaturation force on temperature. The experimental data are taken from \cite{rouzina}. The values of the material parameters are given in the text and are consistent with those reported in \cite{rouzina,peyrardreview,peyrard2}.  Inset: comparison of the potential energy in Eq. (\ref{eq:potentialcut}) (continuous line) and the branch of the Morse potential corresponding to the breaking of a link (dashed line) with $a=60\,\mbox{nm$^{-1}$}$ and $D=12.98\,\mbox{pN\,nm}$ (represented only for positive values of the displacement $y$, see discussion in the text).}
\label{fig:confrontdata} 
\end{figure}

To compare the behavior of our system with respect to the experimental one, we choose the material parameters consistently with those reported in \cite{rouzina,peyrardreview,peyrard2}. Specifically, we consider the following values of the parameters: $k_t=272.34\,\mbox{pN}$, $k_e=76.33\,\mbox{pN}$, $l=0.34\,\mbox{nm}$ and $u_d=0.0145\,\mbox{nm}$. We have $k_t/l=800\,\mbox{pN/nm}=0.05\,\mbox{eV/\AA$^2$}$. Consequently, the critical temperature is $T_c\simeq 372.2 \,\mbox{K}$.  The values of $k_e$ and $u_d$ are fixed to be consistent with the values of the parameters in the Morse potential used in some models of DNA (see for instance \cite{Pey,peyrardreview,peyrard2}). 
Comparing this expression of the Morse potential with the potential energy in Eq. (\ref{eq:potentialcut}) we can impose that $D=k_e\,l/2=12.98\,\mbox{pN\,nm}=0.081\,\mbox{eV}$. In the inset of Figure \ref{fig:confrontdata} we compare the potential energy in Eq.(\ref{eq:potentialcut}) and the Morse potential (with $a=60\,\mbox{nm$^{-1}$}$).  As Figure \ref{fig:confrontdata} shows, the comparison of the experimental data with the theoretical prediction given by Eq.(\ref{eq:fvst}) indicates an excellent agreement. Notwithstanding, the approximations used to obtain explicit analytical formulas (one spatial dimension, extension of the potential energy branches beyond the spinodal point, SDW approximation) does not alter the predictability of the model. 

\begin{figure}
\centering
{\includegraphics[height=5 cm]{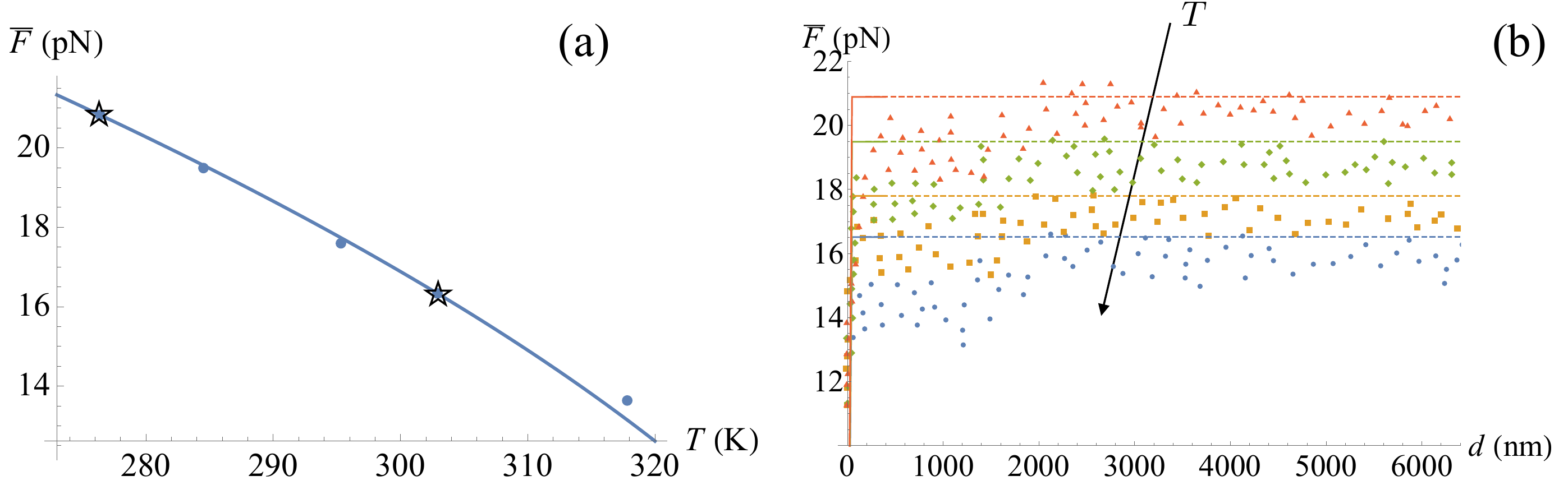}}
\caption{Comparison of experimental results for DNA hairpin from \cite{ritort} with the predictions of the model. (a): unzipping force versus temperature; blue dots are the experimental values whereas the continuous line represents Eq.(\ref{eq:fvst}). (b): Force-extension curves for different temperatures; solid lines are obtained using Eq.(\ref{eq:ansol}) and Eq.(\ref{eq:fdelta}) (rescaling $\bar{f}$ in order to derive the physical force $\bar{F}$ and recalling that $\delta=d/u_d$) whereas dashed lines are obtained from Eq.(\ref{eq:fvst}); from top to bottom the temperatures are $T=276\,\mbox{K}$ (triangles), $T=285\,\mbox{K}$ (diamonds), $T=295\,\mbox{K}$ (boxes), $T=302\,\mbox{K}$ (bullets). The costitutive parameters used to obtain the curves are described in the main text.}
\label{fig:hairpin} 
\end{figure}

Remarkably, due to the generality of the model, our results can be extended to different macromolecules independently from their length and stiffness (assuming the validity of the thermodynamical limit). For low values of base pairs $n$ the discrete results of previous sections can be adopted. As another specific example describing the possibility of wider application of the approach, we can consider the thermomechanical behavior of DNA hairpins as experimentally analyzed in \cite{ritort}. In Figure \ref{fig:hairpin}(a) we report the unzipping force versus the temperature.  In order to fit the experiments, coherently with the previous example, we fix $l=0.34\,\mbox{nm},u_d=0.0165\,\mbox{nm}$ and, using two experimental points marked with a star in \ref{fig:hairpin}(a), we can perform a best fit so to find $k_t=20.03\,\mbox{pN},k_e=108.55\,\mbox{pN}$. Based on these parameters we first can reproduce the results in the force-temperature space. Then, by using the same material paremeters, we predict the force-extension curves at different temperatures as reported in Figure \ref{fig:hairpin}(b). Specifically, we  reproduce the initial part of the curves (solid lines) by using Eq.(\ref{eq:ansol}) and Eq.(\ref{eq:fdelta}) (with $n=6800$ mimicking the number of base pairs in the experiment) whereas we deduce the vaues of the force plateaux using Eq.(\ref{eq:fvst}) (dashed lines). Also in this case, we observe a very good agreement between experimental results and theoretical predictions.

\section{Conclusions}

We considered the thermomechanical action on the melting transition of double stranded DNA molecules. Based on a simplified two-state behavior of the interchains bonds and the introduction of related spin variables, we deduced a full analytic description. Previous analytic results were based on specific assumptions. In particular, in Peyrard et al. \cite{TPM} the authors consider the extreme discretization hypothesis when intra-chains bonds stiffness is negligible as compared with inter-chains, corresponding to small values of $\nu$. In this case the authors obtain a linear dependence on $T$ plus small corrections in $T^2$. This behavior corresponds to the case when entropic terms represent a perturbation of the internal energy terms and the considered configurations can be considered as perturbation of the purely mechanical limit. In particular, in the case of very low temperature, we  deduced  here analytic expressions for all local and global energy minimizers. As we show the themomechanical melting behavior approaches this purely mechanical limit as temperature decreases. In this limit we are also able to describe dissipative effects and hysteresis qualitatively reproducing the cyclic experimental behavior of double stranded DNA molecules that under unloading undergo a rebonding transition with a resulting hysteresis \cite{SCB}. On the other hand, Transfer Integrals approaches (see {\it e.g.} \cite{Singh}) are based on the opposite hypothesis of inter-chains bonds stiffness negligible as compared with intra-chains bonds one. This limit describes the behavior when entropic energy terms prevails. In particular this is the case when the system approaches the denaturation temperature. As we show, our analytical model recover the results obtained here based on the Transfer Integral technique in the limit of large values of $\nu$. Moreover, based on a simplifying assumption on the inter-chains energy (two state, parabolic-constant energy), on the assumption of vertical displacements of the particles, and of single-walls solutions, we are able to deduce a fully analytical solution of the thermomechanical response of double stranded DNA molecules. The comparison with the experimental behavior supports the effectiveness of such assumptions. No specific assumptions on the bonds stiffnesses are considered, so that our results reconcile previous models in the literature for the whole range of $\nu$. This is of interest not only in the case of DNA melting effects, but in more general interactions between two molecules, such as hairpins or proteins domain unfolding. The availability of analytic results can be also of great interest in the perspective of molecular scale design of new materials.

\section{Supporting Information}

Detailed calculations related to equilibrium solutions, global and local energy minima, partition function and temperature effects, thermodynamical limit and critical temperature, Transfer Integral method (PDF).

\begin{acknowledgement}

\vspace{ 0.1 cm}
The authors are grateful for many discussions and deep suggestions of Lev Truskinovsky on the subject of the paper. 

GF and GP have been supported by the Italian Ministry MIUR-PRIN through the project \emph{Mathematics of active materials: From mechanobiology to smart devices} (2017KL4EF3) and by GNFM (INdAM). GF is also supported by INFN through the project QUANTUM, by the FFABR research grant (MIUR) and the PON S.I.ADD. 
\end{acknowledgement}

\bibliographystyle{apalike}

\pagebreak

\clearpage

\section{Supplementary Information}\label{sec:suppmat}
 \renewcommand{\theequation}{SI\arabic{equation}}
\setcounter{page}{1}
\renewcommand{\thefigure}{SI\arabic{figure}}
\setcounter{figure}{0}

\setcounter{equation}{0}

\subsection{Equilibrium solutions}
We consider here the mechanical (zero temperature) limit. In this case the observed configurations correspond to the minima of the energy in Eq. \eqref{hh}.
In order to simplify the analysis of the equilibria, we introduce the tridiagonal matrix
\begin{equation}
\Bz=\nu^2\Lz+\Dz
\end{equation}
where
\begin{equation}
          \Lz=\left [ \ba{ccccccc}
     \displaystyle         2 & - 1   & & & &\mbox{ \large {\bf 0}}\\
              -1& 2 & \ddots & & & \\
                         &    \ddots&\ddots& & & -1 \\
                                       \mbox{ \large {\bf 0}} &   &  & & -1 & 2 \\
              \ea \right ]
\end{equation}
and 
\begin{equation}
          \Dz=\left [ \ba{ccccccc}
     \displaystyle         1-\chi_1&    & & & &\mbox{ \large {\bf 0}}\\
              & 1-\chi_2 &  & & & \\
                         &    &\ddots& & &  \\
                                       \mbox{ \large {\bf 0}} &   &  & &  & 1-\chi_n\\
              \ea \right ].
\end{equation}
%
%

After introducing the vector $\iz_n=\{0,\dots,1\}$ of the canonical base in $\mathbb{R}^{n}$ we can rewrite the energy in the compact form
\begin{equation}\label{eq:nphi}
n\, \phi=\frac{1}{2}\left (\Bz \wz \cdot \wz+\bchi \cdot \bchi-2  \delta \nu^2  \wz \cdot \iz_n  +  \delta ^2\nu^2  \right ),
\end{equation}
where we introduced the $n$-components vector $\wz=\{w_1,w_2,\dots,w_n\}$ and defied the vector $\bchi=\{\chi_1,\dots,\chi_n\}$ of internal variables assigning the phase configuration $\chi_i$ as in Eq. \ref{chi} and explicitly used the constraint of assigned displacement on the final oscillator, i.e. $w_{n+1}=\delta$ and $w_0=0$. We notice that the quantity $\chi=|\bchi|=\bchi \cdot \bchi$ counts the number of broken links in the considered configuration. In order to obtain the relation between the applied displacement $\delta$ and the force experienced by the system, we have to introduce the Lagrange multiplier $f$ coupled to the assigned $\delta$ and minimize the function
\begin{equation}\label{eq:ngb}
n\, g=n\, \phi-f \delta.
\end{equation}
The equilibrium conditions on $g$ read 
\begin{eqnarray}
\label{equil1}
 &&\frac{\partial g}{\partial \wz}=\left(\Bz \wz -  \delta\,  \nu^2  \, \iz_n \right)={\bf 0} \Leftrightarrow  \wz =  \delta\, \nu^2  \Bz^{-1}\iz_n,\vspace{0.4 cm}\\ 
 &&\frac{\partial g}{\partial \delta}=\left( - \nu^2 \wz \cdot \iz_n  +  \delta\, \nu^2 -f \right)=0 \Leftrightarrow  \delta = \wz \cdot \iz_n + \frac{f}{\nu^2}.
 \label{equil2}
\end{eqnarray}
Thus, we get the equilibrium displacements at fixed configuration $\bchi$ as
\begin{equation}
\wz =   \frac{\nu^2\, \Bz^{-1}\iz_n}{\hat k(\bchi)}  \,f,  \quad \delta= \frac{f}{\hat k(\bchi)}, 
\label{equilsol} 
\end{equation}
where
\begin{equation}
 \hat k(\bchi)= \nu^2[1- \nu^2 \Bz^{-1}_{n,n}(\bchi)]
 \label{eq:kappahat}
 \end{equation}
represents the total stiffness of the system (at assigned phase configuration $\bchi$).
Substituting these solutions into Eq. (\ref{eq:nphi}) we obtain the equilibrium energy at fixed $\bchi$ as
\begin{equation}
 n\,  \phi=\frac{  \hat k(\bchi) \delta^2}{2}+\frac{n-p}{2},
\label{equilen} \end{equation}
where, from the definition of $\bchi$, we have that $n-p=\bchi \cdot \bchi$ represents the number of detached links, with $p=0,...,n$.
We also observe that the solution of the equilibrium equations requires the evaluation
of the inverse matrix $\Bz^{-1}$ of the (Hessian) tridiagonal matrix
$\Bz$. It is easy to verify that the matrix $\Bz$ is
always positive definite, so that it is invertible and all the solutions of
(\ref{equil1})-(\ref{equil2}) are local minima of the energy. General iterative formulas \cite{HC,Nab} to
find $\Bz^{-1}$ can be used to show that all the
elements of the inverse matrix are positive definite and decay as the distance from the
diagonal grows. As a consequence, the equilibrium solutions $w_i$ in Eq.(\ref{equilsol}) are monotonic for $f>0$, {\it i.e.} $w_{i+1}\ge w_{i}$ for $i=0\dots,n-1$.
We remark that these solutions must respect the compatibility condition that
all the links with $\chi_i=0$ (respectively $\chi_i=1$) are in the
attached (detached) configuration, i.e. verify $w_i<1$ (respectively $w_i\geq
1$). In particular, this condition, together with the monotonicity result
described above, implies that all the stable equilibrium solutions are
characterized by an initial connected segment of $p$ attached links and $n-p$ detached links, denoted as {\it single domain wall} (SDW) solutions. Thus, we may parametrize each  state by the single parameter $p$ assigning the total number of attached links. 

\subsection{SDW equilibrium solutions}\label{rem:diblock} 

In order to obtain explicit analytical results without resorting to iterative
formulas for the inversion of tridiagonal matrices, we can follow the approach in
\cite{MPPT}. We decompose the equilibrium equations (\ref{equil1}) into two different
equilibrium problems involving the first attached $p$ links and the remaining
detached $n-p$ links. In terms of the configuration vector, we can consider $\bchi=(0,\dots,0,1,\dots,1)$ where the first $p$ elements have $\chi_i=0$ (attached bases) and the remaining $n-p$ are described by $\chi_i=1$ (detached bases). Thus, the first $p$ equations of \eqref{equil1} after easy manipulations  can be rewritten as
 \be \left[ \ba{ccccccc}
    \!\!  \alpha& \!\!  -1& \!\! &\!\! &\!\! & \!\!&\!\!\mbox{ \large {\bf 0}}\\
  \!\!   -1& \!\!\alpha& \!\!  -1&\!\! & \!\!&\!\! &\!\! \\
  \!\!  &  \!\! & \!\!\ddots&\!\!\ddots&\!\!\ddots&\!\! &\!\! \\
    \!\!& \!\!& \!\!  &\!\! &\!\! -1& \!\!\alpha& \!\!  -1\\
   \!\! \mbox{ \large {\bf 0}}&\!\! &\!\!   &\!\!  &\!\! &\!\! -1& \!\!\alpha \\
    \ea\! \right ]\!\!
    \left [ \ba{c}  w_1 \\
    w_2 \\ \vdots \\ w_{p-1} \\ w_{p} \ea \right ]\!\! =\!\!
    \left [  \ba{c}  0\\
    0 \\ \vdots \\ 0 \\ w_{p+1} \ea \right ] \,
 \label{det} \ee
 where $\alpha=2+\frac{1}{\nu^2}$. By using the results in \cite{HC,determinant} we can evaluate the inverse of the tridiagonal matrix in (\ref{det}) and derive the displacements of the $p$ attached elements as
 \begin{equation}
 w_i=\frac{\sinh{(i\lambda)}}{\sinh{[(p+1)\lambda]}}\, w_{p+1}, \,\,\,\, i=1,...,p,
  \end{equation}
where
 \begin{equation}\label{eq:lambda}
 \cosh{\lambda}=1+\frac{1}{2 \nu^2}.
 \end{equation} 
On the other hand, when the links are detached the equilibrium solutions trivially give
 \begin{equation}
 w_{i+1}-w_i=\frac{f}{\nu^2}, \,\,\,\, i=p,...,n.
  \end{equation}
 One then easily deduces
 \begin{equation}\label{bcbc}
w_{p+1}=\alpha(p) \frac{f}{\nu^2},
  \end{equation}
  where
 \begin{equation}
 \alpha(p)=\left \{1-\frac{\sinh{(p \lambda)}}{\sinh{[(p+1)\lambda]}}\right\}^{-1}.
 \end{equation}
  As a result, the equilibrium solutions of Eq.(\ref{equilsol}) for fixed value of unbroken links $p$ (with $p=0,...,n$) can be rewritten as
\begin{equation}
   \left \{\begin{array}{l}
  w_i=\frac{\sinh{(i\lambda)}}{\sinh{[(p+1)\lambda)]-\sinh{(p\lambda)}}} \, \frac{f}{\nu^2 }, \quad i=1,...,p \\

  w_i=\left (\frac{\nu^2}{k(p)}-(n+1-i)\right)\frac{f}{\nu^2 }, \quad i=p+1,...,n
    \end{array}\right.
    \end{equation}
  and 
  \begin{equation}
  w_{n+1}=\delta=\frac{f}{k(p)}.
   \end{equation}
  In these results we introduced the global elastic modulus
 \begin{equation} \label{19}
k(p)=\frac{\nu^2}{\gamma(p)},  \,\,\, \,\,\,\, p=0,\dots,n, 
\end{equation}
where we defined
\begin{equation}
\gamma(p)=n-p+\alpha(p),\quad \alpha(p)=\left \{1-\frac{\sinh{(p \lambda)}}{\sinh{[(p+1)\lambda]}}\right\}^{-1}
\end{equation}
and used Eq.(\ref{eq:kappahat}) noticing that
\begin{equation}\label{eq:Bminus1}
1-\nu^2\Bz^{-1}_{n,n}(p)=\frac{1}{\gamma(p)}.
\end{equation}
Finally, by using \eqref{bcbc} the domain of existence of the $p$-th equilibrium branch can be characterized in terms of $f$ with 
\begin{equation} \label{fmM}
f\in\left (f_{\mbox{\tiny{\it m}}}(p),f_{\mbox{\tiny{\it M}}}(p)\right )= \left ( \frac{ \nu^2}{\alpha(p)},\frac{ \nu^2}{\alpha(p)-1}\right ),
\end{equation}
or in terms of $\delta$ with
\begin{eqnarray}\delta \in \left (\delta_{\mbox{\tiny{\it m}}}(p),\delta_{\mbox{\tiny{\it M}}}(p)\right )= \left ( \frac{ \nu^2 }{\alpha(p)\, k(p)},\frac{ \nu^2}{(\alpha(p)-1)\, k(p)}\right )=\left (1+ \frac{n-p}{\alpha(p)},1+\frac{n-p+1}{\alpha(p)-1}\right ).\label{deltamM}
\end{eqnarray}

\subsection{Global energy minima: Maxwell convention}
\begin{figure}[h!]\vspace{0.8 cm}
\includegraphics[height=6 cm] {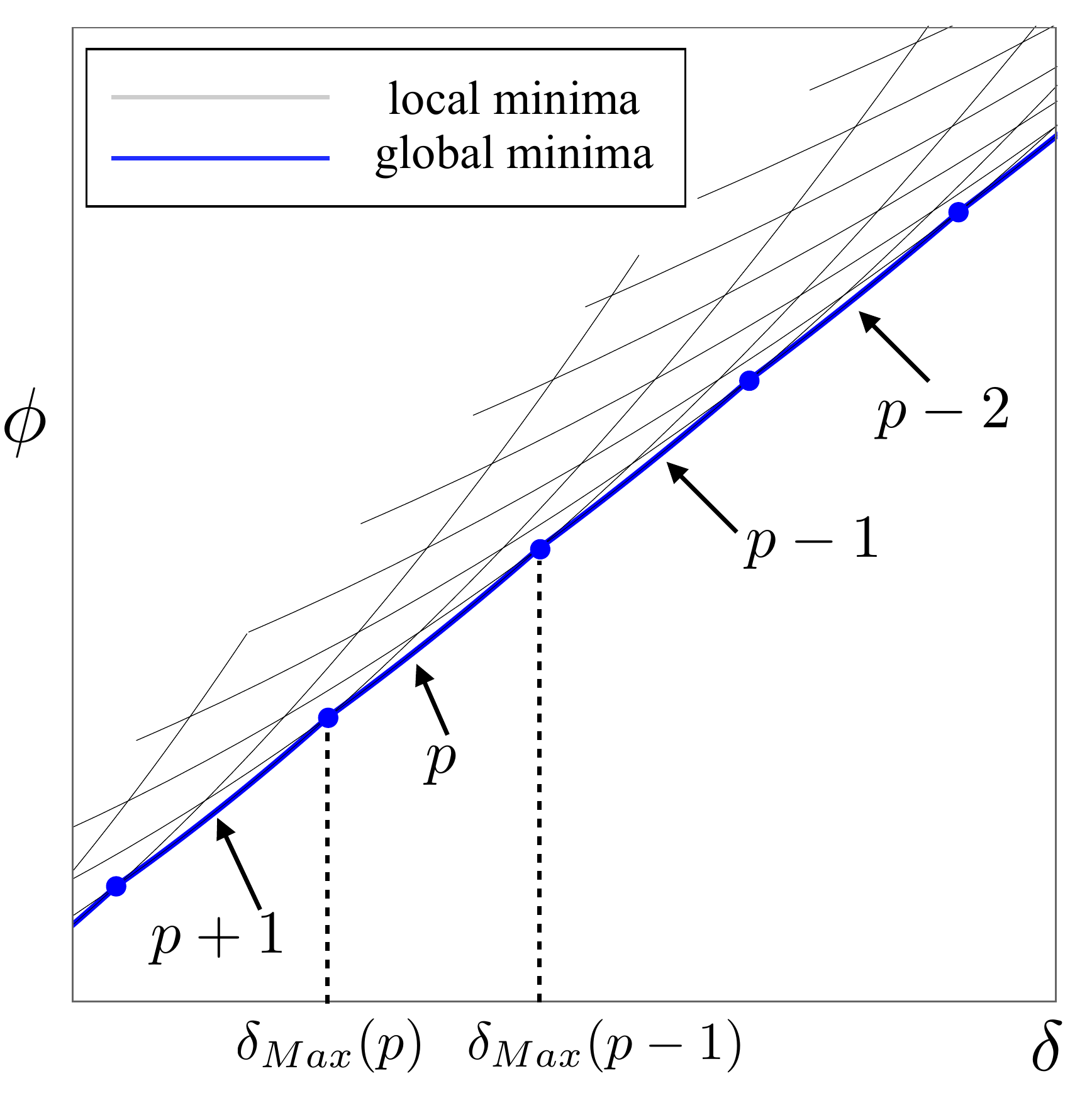}
\caption{Example of energy diagrams. Thin black lines represent local minima, thick blue line global minima. 
 \label{ff2}}
\end{figure}

We consider first the hypothesis that the system relaxes to the global energy minimum as the external assigned displacement is applied and varied. Due to the positivity of the (Hessian) tridiagonal matrix $\Bz$, all the equilibrium branches of the energy in Eq.\eqref{equilen} (identified by the value of $p$) are convex.
Using the definition of $k(p)$ in Eq.(\ref{kk}), we obtain that the equilibrium branches $p$ and $p+1$ of the energy intersect at a single positive value of the assigned displacement given by
\begin{equation}\label{eq:deltamax}
\delta_{Max}(p)=\frac{1}{\nu} \sqrt{\frac{\gamma(p) \, \gamma(p+1)}{\gamma(p)-\gamma(p+1)}}, \,\,\,\, p=0,...,n-1.
\end{equation}
In order to show the behaviour of $\delta_{Max}$ depending on $p$ we observe that
\begin{equation}
\frac{d \gamma(p)}{dp}=\frac{\lambda  \sinh (\lambda )}{[\sinh (p\lambda  )-\sinh (   (p+1)\lambda )]^2}-1,
\end{equation}
and 
\begin{equation}
\lim_{p\rightarrow 0}\frac{d \gamma(p)}{dp}=\frac{\lambda}{\mbox{sinh} \lambda }-1<0,
\end{equation}
for all $\lambda>0$. Moreover, we have that
\begin{equation}
\frac{d^2 \gamma(p)}{dp^2}=\frac{2 \lambda^2  \sinh (\lambda ) [  \cosh ((p+1)\lambda)- \cosh (p\lambda))}{[\sinh (p\lambda)-\sinh ((p+1)\lambda)]^3}<0,
\end{equation}
for all $p>0$. As a consequence, $\gamma$ is a concave decreasing function of $p$ and from Eq.(\ref{eq:deltamax}) we can deduce  that $\delta_{Max}(p)$ decreases for increasing values of $p$. i.e.
\begin{equation}
\frac{d \delta_{\mbox{\tiny Max}}(p)}{dp}<0.
\end{equation}
In the previous calculations we used a slight abuse of notation: $p$ is a discrete variable, but the results have been obtained by considering it as continuous quantity. Using Eq.(\ref{eq:deltamax}) and Eq.(\ref{equilenp}) it is possible to identify the global energy minimum for an assigned displacement $\delta$. One finds that the equilibrium branch labelled by $p$ represents the global energy minimum for
  \begin{equation}\label{eq:deltagm}
  \left \{ \begin{array}{lllll} 
\delta <\delta_{Max}(n-1) & \quad\mbox{if} & &p=n &\quad \mbox{(fully attached solution)} \\
 \delta \in[\delta_{Max}(p),\delta_{Max}(p-1)] & \quad \mbox{if} & & 1\le p\le n-1  & \quad\mbox{(partially detached solution)}  \\
\delta>\delta_{Max}(0) &  \quad\mbox{if} & & p=0  & \quad\mbox{(fully detached solution)}\\ 
 \end{array}\right.
 \end{equation}
In Fig.~\ref{ff2} we show an example of this behavior with the different equilibrium branches of the energy (each corresponding to a different value of $p$) intersecting. Following the Maxwell convention, we have marked the piecewise curve corresponding to the global minimum of the energy.

According to the Maxwell convention, the system behaves elastically with $f=k(n)\delta$ until $\delta=\delta_{Max}(n-1)$. After this value one observes a sequence of jumps between neighbor branches (each denoted by the value of unbroken links $p$). The corresponding measured force $f$ assumes values in a $p$-dependent interval ($p>0$) as $f\in[f_1(p),f_2(p)]$ where $f_1(p)= k(p)\delta_{Max}(p),\, f_2(p)=k(p)\delta_{Max}(p-1)$.

\begin{figure}[t]\vspace{0.2 cm}
\includegraphics[height=6 cm]{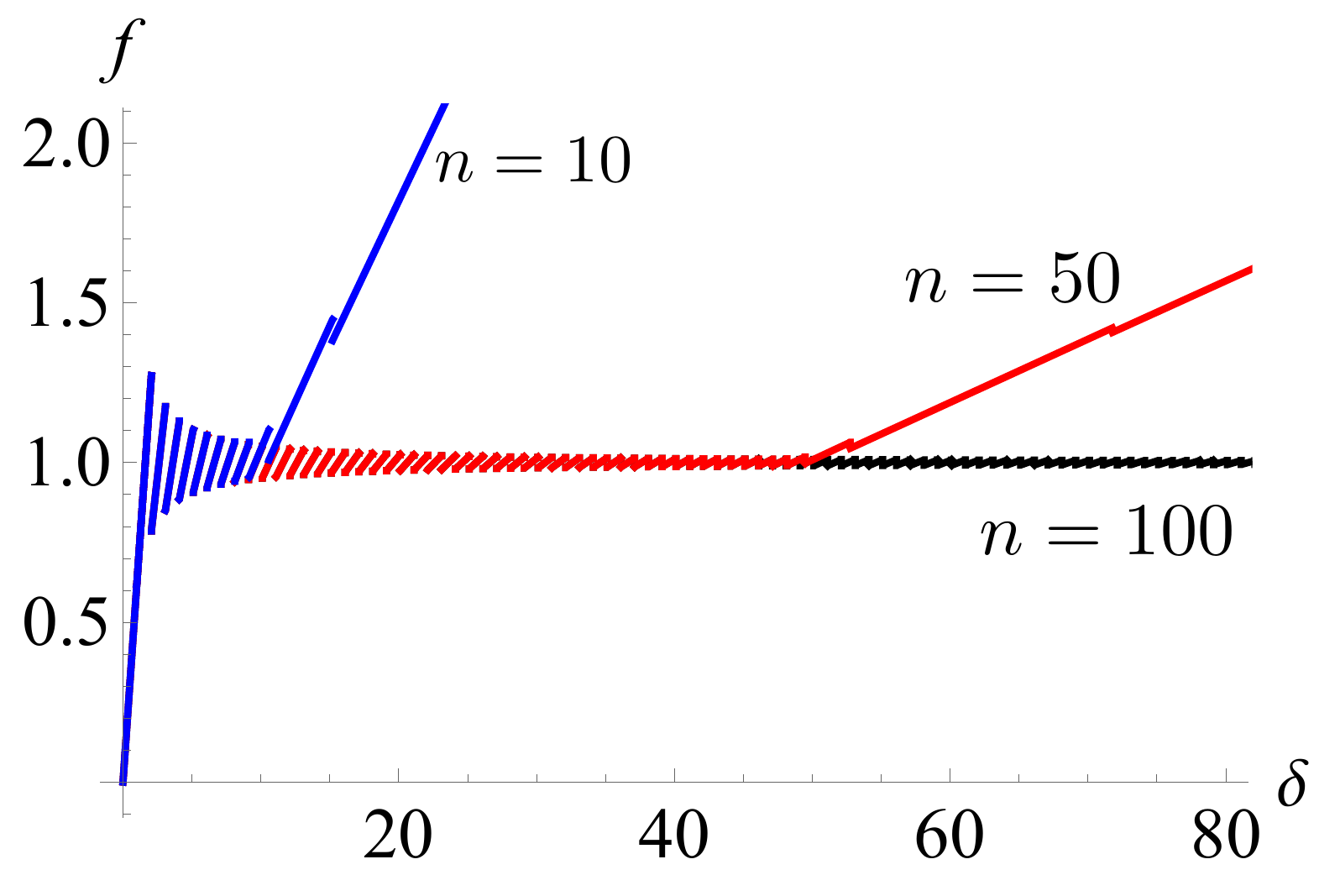}
\caption{Force-displacement relations at for different values of $n$ bonds under the global energy minimization hypothesis.\label{dtl}}
\end{figure}

\paragraph{Thermodynamical Limit}

We now deduce the behavior in the thermodynamic limit $n\rightarrow \infty$ (keeping $l$ fixed). An example of the dependence on the number $n$ of elements is represented in Fig.\ref{dtl}.  

Two different cases can be considered. In the first case we consider the behavior of the system when the breaking of the links start to occur, corresponding to finite values of $r$, where we define $r:=n-p$ as the number of detached elements. An explicit calculation shows that 
\begin{equation}
\begin{array} {l}
\displaystyle \lim_{n\rightarrow\infty}f_{2}(r)= \sqrt{\frac{\left(e^{\lambda }-1\right)
   r+e^{\lambda}}{\left(e^{\lambda }-1\right)  r+1}},\, \nu\vspace{0.2 cm}\\
\displaystyle\lim_{n\rightarrow\infty}f_{1}(r)=  \sqrt{\frac{\left(e^{\lambda }-1\right)
   r+1}{\left(e^{\lambda }-1\right) r+e^{\lambda }}} \, \nu. \vspace{0.2 cm}\\
   \end{array}
\end{equation}
In the second case, we consider a finite fraction of detached elements with $r=\rho n \rightarrow \infty$. The threshold forces assume a particularly simple and clear expression
\begin{equation} 
\lim_{n\rightarrow\infty}f_{2}(\rho n)=\lim_{n\rightarrow\infty}f_{1}(\rho n)=\nu.\label{spsp}
\end{equation}
Observe that these forces do not depend on the fraction of unbroken bonds $\pi=p/n$. 
Moreover, In this limit we have that the equilibrium forces are given by
\begin{equation} f=\frac{\nu^2}{\rho}\frac{\delta}{n},\label{eq:ftl}\end{equation}
with a non zero stiffness. Thus we have that due to the rapid oscillations the convergence of the equilibrium branches is only weak and the stress plateaux  in \eqref{spsp} corresponds to a non zero stiffness of the system that decreases as the percentage $\rho$ of broken links increases.

\subsection{Local energy minima: maximum delay convention}
Based on Eq. \eqref{fmM} we may observe that the breaking force $f_{\mbox{\tiny{\it M}}}(p)$ grows as the number of attached links $p$ decreases. This can be shown by evaluating $f_{\mbox{\tiny{\it M}}}(p)/dp$ and observing that
\begin{equation}
\frac{d\alpha(p)}{dp}=\frac{\lambda  \sinh (\lambda )}{\left[\sinh (\lambda  p)-\sinh (\lambda  (p+1))\right]^2}>0.
\end{equation}
On the other hand, this effect appears numerically negligible.

\section{Temperature effect}

\subsection{Partition function and temperature dependent force-displacement relation}
The partition function is defined as
\begin{eqnarray}
\mathcal{Z}&=&\sum_{\{\chi_i\}}\int_{\R^{n+1}}  e^{-\beta n \phi(\wz,w_{n+1})}\delta_D(w_{n+1}-\delta)\,\, d \wz dw_{n+1}=\sum_{\{\chi_i\}}  \int_{\R^{n}}  e^{-\beta n \phi(\wz,\delta)}\,\, d \wz 
\end{eqnarray}
where we have introduced the constraint $w_0=0$ whereas the Dirac delta distribution $\delta_D$ is used to impose the assigned displacement $w_{n+1}=\delta$. The symbol $\{\chi_i\}$ represent the summation over all configurations with $\chi_i=\{0,1\}$. We defined the rescaled (non dimensional) inverse temperature $\beta$ as
\begin{equation}
\beta=\frac{k_e l}{k_B T}
\end{equation}
where $k_B$ is the Boltzman constant and $T$ the absolute temperature.
The explicit calculation, based on a Gaussian integration, gives
\begin{equation}\label{eqn:ZN1delta}
\mathcal{Z}=\left(\frac{2 \pi}{\beta}\right)^{n/2}\sum_{\{\chi_i\}} \Gamma_{\mbox{{\tiny \bchi}}}(\beta, \delta),
\end{equation}
where 
\begin{equation}\label{eq:Gamma}
\Gamma_{\mbox{{\tiny \bchi}} }(\beta, \delta)= \frac{1}{\sqrt{\mbox{\footnotesize det} \Bz}} e^{-\frac{\beta}{2} \chi} e^{-\frac{\beta}{2}\hat{k}({\mbox{{\tiny \bchi}}}) \delta^2}
\end{equation}
and the total stiffness 
$\hat{k}({\mbox{{\tiny \bchi}}})$ is defined in Eq. (\ref{eq:kappahat}).

\begin{figure}[t!]
\centering
{\includegraphics[height=11 cm]{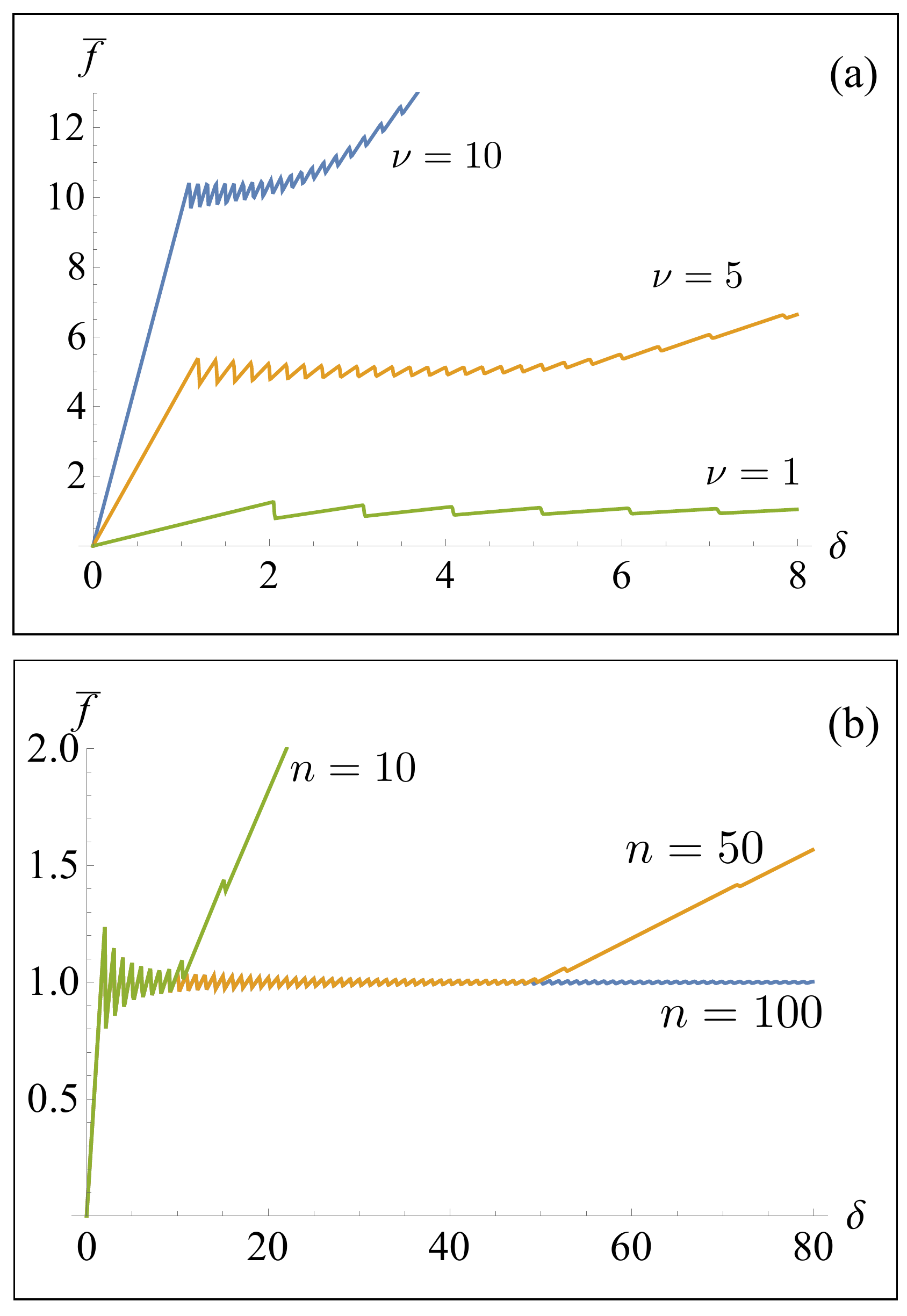}}
\caption{(a) Force-displacement curves for different values of $\nu$ ($n=30$, $\beta=10^3$). (b) Force-displacement curves for different values of $n$ ($\nu=1$, $\beta=10^3$).}
\label{fig:hardconfrontnunumber} 
\end{figure}

The free energy associated to the partition function $\mathcal{Z}$ is
\begin{equation}\label{GG}
\mathcal{F}(\beta, \delta)=-\frac{1}{\beta}\ln \mathcal{Z}(\beta, \delta)=-\frac{n}{2\beta}\ln\frac{2\pi}{\beta}-\frac{1}{\beta}\ln\sum_{\{\chi_i\}}\Gamma_{\mbox{{\tiny \bchi}} }(\beta, \delta).
\end{equation}
The expectation value of the equilibrium force $\langle f\rangle=\bar{f}$ conjugated to $\delta$ can be evaluated by using $
\frac{\partial \mathcal{F}}{\partial\delta}=\bar{f},$ that gives 
\begin{equation}\label{eq:fdelta}
\langle f(\beta, \delta)\rangle=\bar{f}(\beta, \delta)=\frac{\partial \mathcal{F}}{\partial\delta}= \displaystyle \frac{\sum_{\{\chi_i\}}\hat{k}(\bchi)\Gamma_{\mbox{{\tiny \bchi}}}(\beta, \delta)}{\sum_{\{\chi_i\}}\Gamma_{\mbox{{\tiny \bchi}}}(\beta, \delta)} \, \delta.
\end{equation}
This expression describes the relation between the applied displacement and the force experienced by the last element of the chain representing the DNA in our model. In particular, we notice that the expectation value of the force is a statistical superposition of the different configurations with the statistical weight depending on temperature and $\delta$. On the other hand, Eq.(\ref{eq:fdelta}) is difficult to be handled. For instance, we notice that the summation over all configurations of $\bchi$ in the partition function scales exponentially with $n$.  In order to simplify the analysis of the force-displacement relation and obtain fully analytical results we resort to the SDW solutions (with SDW equilibrium solutions corresponding to the global energy minima) already introduced in the mechanical (zero temperature) case. When temperature effects are included, considering only this class of configurations is an approximation. On the other hand,  this approach is justified in the temperature range of real systems\cite{FPFP}. In particular, we have
\begin{equation}\label{eq:fdeltasdw}
\bar{f}(\beta, \delta)\simeq\displaystyle \frac{\sum_{p=0}^n k(p)\Gamma_{p}(\beta, \delta)}{\sum_{p=0}^n \Gamma_{p}(\beta, \delta)} \, \delta,
\end{equation}
where $p$ represents the number of unbroken links and
 \begin{equation}\label{eq:Gammasdw}
\Gamma_{p}(\beta, \delta)= \frac{1}{\sqrt{\mbox{\footnotesize det} \Bz(p)}} e^{-\frac{\beta}{2} (n-p)} e^{-\frac{\beta}{2}k(p) \delta^2}.
\end{equation}
In Fig.~\ref{fig:hardconfrontnunumber} (a)-(b) we plot the force-displacement curves for different values of $\nu$ and $n$, respectively. The results (here shown for $\beta=10^3$) are consistent with those that can obtained in the mechanical (zero temperature) limit. 

It is possible to obtain the expectation value of each rescaled displacement $w_i$ depending on $\delta$ and $T$. An explicit calculation based on Gaussian integration gives
\begin{equation}\label{eq:wi}
\langle w_i(\beta, \delta)\rangle=\bar{w}_i(\beta, \delta)=\displaystyle \frac{\sum_{\{\chi_i\}}\Bz^{-1}_{i,n}(\bchi)\Gamma_{\mbox{{\tiny \bchi}}}(\beta, \delta)}{\sum_{\{\chi_i\}}\Gamma_{\mbox{{\tiny \bchi}}}(\beta, \delta)} \, \nu^2\,\delta.
\end{equation}
In the SDW approximation we have
\begin{equation}\label{eq:wisdw}
\bar{w}_i(\beta, \delta)\simeq\displaystyle \frac{\sum_{p=0}^n \Bz^{-1}_{i,n}(p)\Gamma_{p}(\beta, \delta)}{\sum_{p=0}^n \Gamma_{p}(\beta, \delta)} \,  \nu^2\,\delta.
\end{equation}

\subsection{Thermodynamic limit: saddle point approximation and critical temperature}

\begin{figure}
\centering
{\includegraphics[height=6 cm]{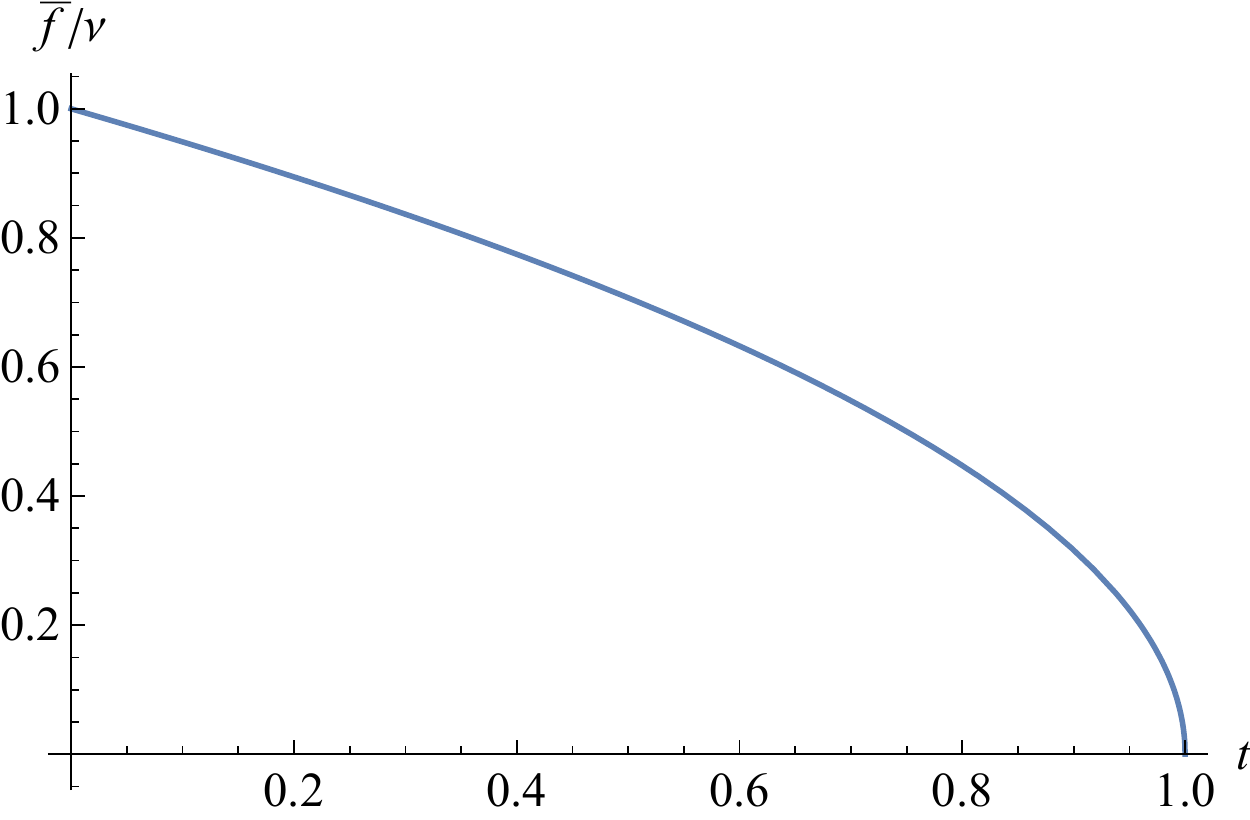}}
\caption{Rescaled average force $\bar{f}/\nu$ with respect to $t=T/T_c$.}
\label{fig:tcritical} 
\end{figure}

Let us consider Eqs.(\ref{eq:fdeltasdw})-(\ref{eq:Gammasdw}). In order to evaluate these expression in the limit of large $n$, we need to compute $\mbox{det} \Bz$ and $k(p)$ from Eq.(\ref{kk}). Using formulas for tridiagonal matrices it is possible to show that
\begin{eqnarray}\label{eq:detBdiblock}
\mbox{\footnotesize det} \Bz&=&\frac{\nu^{2n}}{\sinh\lambda}\left[(n-p+1)\sinh[(p+1)\lambda]-(n-p)\sinh(p\lambda)\right]
\end{eqnarray}
Using this result and introducing the fraction of unbroken links $\pi=p/n\ne0$, we have 
\begin{equation}
\begin{array}{l}\label{eq:detBtdlimit}
\frac{1}{\sqrt{\mbox{\footnotesize det} \Bz(p)}}=\sqrt{\frac{\sinh\lambda}{n\nu^{2n}}}\left[\left(\frac{1}{n}+1-\pi\right)\sinh\left(\lambda n(\pi+1/n)\right)-\left(1-\pi\right)\sinh\left(\lambda n \pi)\right)\right]^{-1/2}\\ 
=\sqrt{\frac{2\sinh\lambda}{n\nu^{2n}}}\left[e^{n\lambda (\pi+1/n)}\left((1-e^{-\lambda})(1-\pi)+\frac{1}{n}\right)+e^{-n\lambda\left(\pi+1/n\right)}\left((e^\lambda-1)(1-\pi)-\frac{1}{n}\right)\right]^{-1/2}  \\
\simeq\sqrt{\frac{2\sinh\lambda}{n\nu^{2n}(1-e^{-\lambda})}}\frac{1}{\sqrt{(1-\pi)}}e^{-\frac{1}{2}n\lambda \pi},
\end{array}
\end{equation}
where in the final passage we have obtained the asymptotic formula for $n\rightarrow \infty$. Moreover, recalling Eq.(\ref{19}), in the $n\rightarrow \infty$ limit we have
\begin{equation}\label{eq:ktdlimit}
n k(n\pi)\rightarrow \frac{\nu^2}{1-\pi}.
\end{equation}
Following Eq.(\ref{eq:Gammasdw}), we have for $n\rightarrow \infty$
\begin{equation}
e^{-\frac{\beta}{2} (n-p)} e^{-\frac{\beta}{2}k(p) \delta^2}\simeq e^{-n\frac{\beta}{2} \left[1-\pi+\frac{\nu^2}{1-\pi} \left(\frac{\delta}{n}\right)^2\right]}.
\end{equation}
Combining these results into Eq.(\ref{eq:fdeltasdw}) we have that the expectation value of the force corresponding to an assigned rescaled displacement $\tilde{\delta}=\delta/n$ in the large $n$ limit is given by 
\begin{eqnarray}\label{eq:fnlarge1}
\bar{f}\simeq\frac{\int_0^1\frac{\nu^2}{1-\pi}\frac{1}{\sqrt{1-\pi}}e^{-n\eta(\pi,\tilde{\delta})}d\pi}{\int_0^1 \frac{1}{\sqrt{1-\pi}}e^{-n\eta(\pi,\tilde{\delta})}d\pi}\tilde{\delta},
\end{eqnarray}
where the sum over $p=0,...,n$ has been substituted by the integral over $\pi=p/n$ and we defined
\begin{equation}
\eta(\pi,\tilde{\delta})=\frac{1}{2}\beta \left(1-\pi+\frac{\nu^2}{1-\pi} \tilde{\delta}^2\right)+\frac{1}{2}\lambda \pi.
\end{equation}
For $n$ large we can use the saddle point approximation, solve the equation $\partial\eta/\partial\pi=0$ and find the critical points
\begin{equation}
\pi_{\pm}=1\pm\nu\tilde{\delta}\sqrt{\frac{\beta}{\beta-\lambda}}.
\end{equation}
Due to the positivity of $\pi$, we consider the solution 
\begin{equation}\label{eq:pistar}
\pi^*=\pi_{-}=1-\nu\tilde{\delta}\sqrt{\frac{\beta}{\beta-\lambda}}.
\end{equation}
We can approximate the integrals in Eq.(\ref{eq:fnlarge1}) to obtain 
\begin{eqnarray}\label{eq:ftdlimit}
\bar{f}&\simeq&\frac{\nu^2 \tilde{\delta}}{1-\pi^*}\frac{\int_0^1\frac{1}{\sqrt{1-\pi^*}}e^{-n\left[\eta(\pi^*,\tilde{\delta})+\frac{1}{2}\eta''(\pi^*,\tilde{\delta})(\pi-\pi^*)\right]}d\pi}{\int_0^1 \frac{1}{\sqrt{1-\pi^*}}e^{-n\left[g(\pi^*,\tilde{\delta})+\frac{1}{2}\eta''(\pi^*,\tilde{\delta})(\pi-\pi^*)\right]}d\pi}=\nu\sqrt{1-\frac{\lambda k_B T}{k_e l}}=\nu\sqrt{1-\frac{T}{T_c}},
\end{eqnarray}
where we have defined the critical temperature (independent on the applied displacement) 
\begin{eqnarray}
T_c&=&\frac{k_e \,l}{k_B\,\lambda }=\frac{k_e\, l}{ k_B \,\mbox{arccosh}(1+\frac{1}{2\nu^2})}=\frac{k_e\, l}{ k_B\, \mbox{arccosh}\left(1+\frac{k_e\,l^2}{2k_t\,u_d^2}\right)}.
\end{eqnarray} 
In Figure \ref{fig:tcritical} we plot the dependence of the rescaled average force $\bar{f}/\nu$ with respect to the adimensional parameter $t=T/T_c$.

The dependence of the number of detached elements $r=n-p$ on temperature and applied displacement can be obtained. The expectation value of the fraction of detached elements (in the SDW approximation) reads
\begin{eqnarray}
\frac{\langle r\rangle}{n}=\langle \rho\rangle=\frac{1}{\mathcal{Z}}\sum_{p=0}^n \int_{\R^{n}}\left(1-\frac{p}{n}\right) e^{-\beta  n \phi(\wz,\delta)} d \wz.
\end{eqnarray}
A calculation analogous to the one used to evaluate the partition function gives
\begin{eqnarray}
\langle \rho\rangle=\frac{\sum_{p=0}^n \left(1-\frac{p}{n}\right)\Gamma_p(\beta, \delta)}{\sum_{p=0}^n \Gamma_p(\beta, \delta)}
\end{eqnarray}
where the definition of $\Gamma_p$ is given in Eq. (\ref{eq:Gammasdw}).
In the thermodynamic limit $n\rightarrow \infty$ we can use again the saddle point approximation and obtain
\begin{eqnarray}
\langle \rho\rangle&\simeq&\frac{\int_0^1{(1-\pi)}\frac{1}{\sqrt{1-\pi}}e^{-n\eta(\pi,\tilde{\delta})}d\pi}{\int_0^1 \frac{1}{\sqrt{1-\pi}}e^{-n\eta(\pi,\tilde{\delta})}d\pi}\simeq 1-\pi^*=\nu\tilde{\delta}\sqrt{\frac{T_c}{T_c-T}}.
\end{eqnarray}
Thus, the number of detached elements is 
\begin{eqnarray}\label{eq:detacH}
\langle r\rangle=\langle n-p\rangle\simeq\nu\delta\sqrt{\frac{T_c}{T_c-T}}.
\end{eqnarray}
When $T\rightarrow T_c$ we notice that $\langle r\rangle$ diverges for any finite value of the applied displacement $\delta$. This is coherent with previous interpretation of $T_c$ as  the denaturation temperature. Obviously, the fraction of attached bonds $\pi$ in the thermodynamic limit  is given by $\pi^*$ (using the saddle point method). In particular, it is possible to search the value of applied displacement $\tilde{\delta}_c$ such that one observes the complete thermally induced denaturation and the fraction of attached bonds is zero. From Eq.(\ref{eq:pistar}) we solve the equation $\pi^*=0$ and find
\begin{equation}
\tilde{\delta}_c=\frac{1}{\nu}\sqrt{1-\frac{T}{T_c}}.
\end{equation}

\subsection{Thermodynamic limit: Transfer Integral method}

Using Eq.(\ref{eq:enelastic}), (\ref{eq:potentialcut}) and (\ref{eq:nu}) we rewrite the partition function of the system as 
\begin{eqnarray}\label{eq:partitionTI}
\mathcal{Z}&=&\int_{\R^{n+2}} \prod_{i=0}^n dw_i\,\delta_D(w_0)\,dw_{n+1} e^{-\beta \frac{\nu^2}{2}(w_1-w_0)^2}\left\{\prod_{j=1}^{n-1}e^{-\beta \left[\frac{\nu^2}{2}(w_{j+1}-w_j)^2+V(w_j)\right]}\right\}\nonumber\\
&\times&e^{-\beta \left[\frac{\nu^2}{2}(w_{n+1}-w_n)^2+V(w_n)\right]}\delta_D(w_{n+1}-\delta),
\end{eqnarray}
where
\begin{equation}\label{eq:V}
V(y)=\frac{1}{2} \left \{ \begin{array}{ll} y^2,  &\mbox{ if } |y|\leq 1 \\
\\
1  &\mbox{ if } |y|>1 \end{array}\right .
\end{equation}
Following \cite{Pey}, let us define the Transfer Integral (TI) operator and its eigenfunctions $\psi$ as
\begin{equation}\label{eq:TI}
 \int dx e^{-\beta\left[\frac{\nu^2}{2}(y-x)^2+V(y)\right] }
\psi(x)  = e^{-\beta \epsilon} \psi(y).
\end{equation}
Moreover, we can expand $\delta_D(w_{n+1}-\delta)$ in Eq.(\ref{eq:partitionTI}) in the basis of the eigenfunctions $\psi_i$ as
\begin{equation}
 \delta_D(w_{n+1}-\delta)=\sum_i\psi_i^*(\delta)\psi_i(w_{n+1}).
\end{equation}
The evaluation of the partition function is performed by successive integration over the variables $w_i$ ad using the definition of TI. We obtain
\begin{equation}
\mathcal{Z}=\sum_i\psi_i^*(\delta)\psi_i(0)e^{-\beta(n+1)\epsilon_i}.
\end{equation}
In the thermodynamic limit it is possible to consider only the contribution of the smallest value of $\epsilon_i$ (and the corresponding $\psi_0$), thus obtaining
\begin{equation}\label{eq:ZapproxTI}
\mathcal{Z}_{TI}\simeq\psi_0^*(\delta)\psi_0(0)e^{-\beta(n+1)\epsilon_0}.
\end{equation}
The expectation value of the force can be again obtained from the free energy $\mathcal{F}_{TI}$ and the partition function. We have
\begin{equation}\label{eq:forceTI}
\bar{f}_{TI}=\frac{\partial\mathcal{F}_{TI}}{\partial\delta}=-\frac{1}{\beta\mathcal{Z}_{TI}}\frac{\partial\mathcal{Z}_{TI}}{\partial\delta}
\end{equation}
Thus, in order to obtain the force-displacement relation it is necessary to evaluate the eigenfunctions and the eigenvalues of the TI in Eq.(\ref{eq:TI}). We can derive analytical expressions {\it in the regime} 
\begin{equation}\nu^2\gg 1.\end{equation}
In particular, the main contribution to the integral comes from a region where $x\simeq y$. As a consequence, we can expand $\psi$ in powers of $z=x-y$ up to $O(z^2)$ and evaluate Eq.(\ref{eq:TI}) using a Gaussian integration in $z$. Finally, we arrive at an ordinary differential equation of the form
\begin{equation}\label{eq:partialTI}
\frac{1}{2\beta\nu^2}\frac{d^2\psi(y)}{d y^2} + \psi(y)=e^{-\beta\left[E-V(y)\right]}\psi(y),
\end{equation}
where $E=\epsilon+\frac{1}{2\beta}\ln\frac{2\pi}{\beta\nu^2}$. Supposing that $\beta\left[E-V(y)\right]<1$ we can expand the exponential function on the right-hand side of Eq.(\ref{eq:partialTI}) and obtain the final equation
\begin{equation}\label{eq:partialTIfinal}
-\frac{1}{2(\beta\nu)^2}\frac{d^2\psi(y)}{d y^2} + V(y)\,\psi(y)=E\,\psi(y).
\end{equation}
Formally, we derived a time-independent Schroedinger equation whose solution correspond to study a `barrier penetration problem' with an assigned energy $E$ smaller than the height of the barrier \cite{messiah}. In our case the height is 
\begin{equation}
V_0=|V(\pm1)|=1/2. 
\end{equation}
In principle, this problem can be solved using the so-called WKB approximation. Here we use the spin variable introduced to substitute the potential energy in Eq.(eq:potentialcut) and perform analytical calculations in the mechanical limit and for the partition function. If we extend the harmonic part of the potential beyond the region $|y|\leq 1$ in Eq.(\ref{eq:V}) the Schroedinger equation reduces to the problem of the quantum harmonic oscillator whose eigenvalues and eigenfuntions are known \cite{messiah}. In particular, we have that the ground state energy is
\begin{equation}\label{eq:gsenergy}
E_0=\frac{1}{2\beta\nu}
\end{equation}
and the eigenfunctions are Hermite polynomials.
On the other hand, in the flat potential region $y> 1$ the solution of the Schroedinger equation for $E<V_0$ reads
\begin{equation}\label{eq:solutionflat}
\psi_0(y)\propto e^{-\sqrt{2(\beta\nu)^2(V_0-E_0)}y},
\end{equation}
where we have explicitly introduced $E_0$ in place of $E$ (the values of the energy in the different regions must be the same).
We notice that for $y<1$ the solution is the same, but for a a plus sign at the exponent. 
These solutions are valid if $E_0\le V_0$. This condition allows us to obtain the critical temperature predicted by the TI method within the approximations of our model. In particular, we find
\begin{equation}
E_0\le V_0 \iff \frac{1}{\beta}\le \nu=\frac{1}{\beta_c}.
\end{equation}
Taking into account the definitions of $\beta$ and $\nu$ we obtain the critical temperature in terms of the material parameters of the model:
\begin{equation} \label{eq:TcTI}
T_c^{(TI)}=\frac{\sqrt{k_t\,k_e}}{k_B}u_d.
\end{equation} 
We notice that the critical temperature obtained in Eq.(\ref{eq:TcTI}) can be derived from Eq.(\ref{eq:Tc}) in the regime $\nu^2\gg1$ which is consistent with the approximation used to apply the TI method. Finally, using Eq. (\ref{eq:ZapproxTI}), (\ref{eq:forceTI}), (\ref{eq:gsenergy}) and (\ref{eq:solutionflat}) it is possible to obtain the temperature dependent force needed to obtain the complete denaturation of DNA i.e. the breaking of all the bonds of the chain. We find
\begin{equation}
\bar{f}_{TI}=\nu\sqrt{1-\frac{T}{T_c^{(TI)}}},
\end{equation}
consistent with the result in Eq.(\ref{eq:forcetd}). The comparison with the transfer integral technique shows the coincidence of the results about the stiffness dependent denaturation temperature. 

\bibliographystyle{apalike}

\end{document}